%% file: MAIN.tex
\documentclass{ieeeaccess}
\usepackage{cite}
\usepackage{amsmath}
\usepackage{algorithmic}
\usepackage{graphicx}
\usepackage{textcomp}
\def\BibTeX{{\rm B\kern-.05em{\sc i\kern-.025em b}\kern-.08em
    T\kern-.1667em\lower.7ex\hbox{E}\kern-.125emX}}
\usepackage{fontawesome}
\usepackage[nolist]{acronym}
\usepackage{multirow}
\usepackage{url}
\usepackage{subcaption}
\usepackage{academicons}
\usepackage{fancyhdr}
\usepackage{silence}
\usepackage[normalem]{ulem}

\usepackage[table]{xcolor} 
\usepackage{tabularx} 

\usepackage{dblfloatfix}
\usepackage{rotating}
\usepackage{booktabs}
\usepackage{multirow}
\usepackage{multicol}
\usepackage[table]{xcolor}
\usepackage{longtable}
\usepackage{lscape}
\usepackage{lipsum}
\usepackage{enumitem}
\usepackage[utf8]{inputenc}

\usepackage{fontawesome}

\usepackage[english]{babel}

\usepackage{tabularx,booktabs}
\usepackage{dingbat}
\usepackage{diagbox}
\newcolumntype{C}{>{\centering\arraybackslash}X}
\begin{document}
\history{Date of publication 2023, July.}
\doi{10.1109/ACCESS.2023.3292788}

\title{Enhancing Network Slicing Architectures with Machine Learning, Security, Sustainability and Experimental Networks Integration}
\author{\uppercase{Joberto S. B. Martins}\authorrefmark{1} \IEEEmembership{Life Senior Member, IEEE}, \uppercase{Tereza C. Carvalho}\authorrefmark{2}, \uppercase{Rodrigo Moreira}\authorrefmark{3}, 
 \uppercase{Cristiano Both}\authorrefmark{4}, \uppercase{Adnei Donatti}\authorrefmark{2}, \uppercase{João H. Corrêa}\authorrefmark{5}, \uppercase{José A. Suruagy}\authorrefmark{6} 
\IEEEmembership{Member, IEEE}, \uppercase{Sand L. Corrêa}\authorrefmark{7}, \uppercase{Antonio J. G. Abelem}\authorrefmark{8} \IEEEmembership{Member, IEEE}, \uppercase{Moisés R. N. Ribeiro}\authorrefmark{9}, \uppercase{Jose-Marcos Nogueira}\authorrefmark{10}
\IEEEmembership{Member, IEEE}, \uppercase{Luiz C. S. Magalhães}\authorrefmark{11}, \uppercase{Juliano Wickboldt}\authorrefmark{12}, \uppercase{Tiago Ferreto}\authorrefmark{13} \IEEEmembership{Member, IEEE}, \uppercase{Ricardo Mello}\authorrefmark{9}, \uppercase{Rafael Pasquini}\authorrefmark{14} \IEEEmembership{Senior Member, IEEE}, \uppercase{Marcos Schwarz}\authorrefmark{15}, \uppercase{Leobino N. Sampaio}\authorrefmark{16}\IEEEmembership{Member, IEEE}, \uppercase{Daniel F. Macedo}\authorrefmark{10}, \uppercase{José F. de Rezende}\authorrefmark{17}, \uppercase{Kleber V. Cardoso}\authorrefmark{7}, and \uppercase{Flávio O. Silva}\authorrefmark{14} \IEEEmembership{Member, IEEE}.}

\address[1]{Universidade Salvador (UNIFACS), Salvador, Brazil (e-mail: joberto@ieee.org)}
\address[2]{Universidade de São Paulo (USP), São Paulo, Brazil (e-mail: terezacarvalho@usp.br, adnei.donatti@usp.br.)}
\address[3]{Universidade Federal de Viçosa (UFV), Viçosa, Brazil (e-mail: rodrigo@ufv.br) }
\address[4]{Universidade Federal do Vale dos Sinos (UNISINOS), Porto Alegre, Brazil (e-mail: cbboth@unisinos.br) }
\address[5]{Universidade Federal do Ceará (UFC), Fortaleza, Brazil (e-mail: joaocorrea@ufc.br) }
\address[6]{Universidade Federal de Pernambuco (UFPE), Recife, Brazil (e-mail: suruagy@cin.ufpe.br) }
\address[7]{Universidade Federal de Goiás (UFG), Goiania, Brazil (e-mail: sandluz@ufg.br, kleber@ufg.br) }
\address[8]{Universidade Federal do Pará (UFPA), Belém, Brazil (e-mail: abelem@ufpa.br) }
\address[9]{Universidade Federal do Espírito Santos (UFES), Vitória, Brazil (e-mail: moises@ele.ufes.br, ricardo.c.mello@ufes.br) }
\address[10]{Universidade Federal de Minas Gerais (UFMG), Belo Horizonte, Brazil (e-mail: jmarcos@dcc.ufmg.br, damacedo@dcc.ufmg.br) }
\address[11]{Universidade Federal Fluminense (UFF), Niteroi, Brazil (e-mail: luizschara@id.uff.br) }
\address[12]{Universidade Federal do Rio Grande do Sul (UFRGS), Porto Alegre, Brazil (e-mail: jwickboldt@inf.ufrgs.br) }
\address[13]{Pontifícia Universidade Católica do Rio Grande do Sul (PUCRS), Porto Alegre, Brazil (e-mail: tiago.ferreto@pucrs.br) }
\address[14]{Universidade Federal de Uberlândia (UFU), Uberlândia, Brazil (e-mail: rafael.pasquini@ufu.br, flavio@ufu.br) }
\address[15]{Rede Nacional de Pesquisa (RNP), Rio de Janeiro, Brazil (e-mail: marcos.schwarz@rnp.br) }
\address[16]{Universidade Federal da Bahia (UFBA), Salvador, Brazil (e-mail: leobino@ufba.br) }
\address[17]{Universidade Federal do Rio de Janeiro (UFRJ), Rio de Janeiro, Brazil (e-mail: rezende@land.ufrj.br) }

\tfootnote{This work was supported by the Fundação de Apoio à Pesquisa do Estado de São Paulo (FAPESP) - Research Project 2018/23097-3, National Council for Scientific and Technological Development (CNPq), ANIMA Institute, and CNPq Grant 421944/2021-8.\\\copyright 2023 IEEE.  Personal use of this material is permitted.  Permission from IEEE must be obtained for all other uses, in any current or future media, including reprinting/republishing this material for advertising or promotional purposes, creating new collective works, for resale or redistribution to servers or lists, or reuse of any copyrighted component of this work in other works.}


\corresp{Corresponding author: Joberto S. B. Martins (e-mail: joberto@ieee.org).}

\begin{abstract}
Network Slicing (NS) is an essential technique extensively used in 5G networks computing strategies, mobile edge computing, mobile cloud computing, and verticals like the Internet of Vehicles and industrial IoT, among others. NS is foreseen as one of the leading enablers for 6G futuristic and highly demanding applications since it allows the optimization and customization of scarce and disputed resources among dynamic, demanding clients with highly distinct application requirements. Various standardization organizations, like 3GPP's proposal for new generation networks and state-of-the-art 5G/6G research projects, are proposing new NS architectures. However, new NS architectures have to deal with an extensive range of requirements that inherently result in having NS architecture proposals typically fulfilling the needs of specific sets of domains with commonalities. The Slicing Future Internet Infrastructures (SFI2) architecture proposal explores the gap resulting from the diversity of NS architectures target domains by proposing a new NS reference architecture with a defined focus on integrating experimental networks and enhancing the NS architecture with Machine Learning (ML) native optimizations, energy-efficient slicing, and slicing-tailored security functionalities. The SFI2 architectural main contribution includes the utilization of the slice-as-a-service paradigm for end-to-end orchestration of resources across multi-domains and multi-technology experimental networks. In addition, the SFI2 reference architecture instantiations will enhance the multi-domain and multi-technology integrated experimental network deployment with native ML optimization, energy-efficient aware slicing, and slicing-tailored security functionalities for the practical domain.
\end{abstract}


\markboth
{Martins J. S. B., Carvalho, T. \headeretal: Enhancing Network Slicing Architectures}
{Martins J. S. B., Carvalho, T. \headeretal: Enhancing Network Slicing Architectures}

\titlepgskip=-15pt

\maketitle

\input{Acronym/acronym}

\section{Introduction}
\label{sec:introduction}
\PARstart{N}{etwork} slicing (NS) is being extensively used in domains such as 5G, the Internet of Things (IoT), the Internet of Vehicles (IoV), Industry 4.0, drone networks, smart transportation systems, smart health care, and smart grids, among others \cite{barakabitze_5g_2020, wijethilaka_survey_2021, mlika_network_2021, wu_survey_2022, yuan_airslice_2020}. However, given its large domain of application and inherent requirements, network slicing architectures and deployment aspects like multi-domain coverage, technology-specific proposals, end-to-end effectiveness, lightweight deployments, and network segment integration are still open research issues.

The \ac{SFI2}\footnote{Available at \url{https://sites.google.com/view/sfi2/home}} research project \cite{dias_sfi2_2019} leading technical and innovative emphasis is to develop a network slicing solution that provides orchestration and allocation of resources for multi-domain experimentation network infrastructures. The experimentation testbeds FIBRE-NG \cite{abelem_fit_2013}, FUTEBOL \cite{both_futebol_2019}, CloudNEXT \cite{atmosphere_cloudnext_2019}, FIWARE \cite{oliveira_improving_2018}, 5GINFIRE \cite{silva_5ginfire_2019}, and NECOS \cite{9422346}, are the main targets of the SFI2 project. In this context, the SFI2 project aims to integrate resources and services from the legacy experimental infrastructures mentioned above by defining a reference architecture and functionalities, allowing its instantiation and deployment on distinct domains.

The SFI2 reference architecture proposed in this article is a new enhanced architecture for network slicing and a practical realization of the Slice-as-a-Service (SlaaS) paradigm, with intelligent end-to-end slice orchestration, considering security requirements and sustainability aspects at different stages of the slice life cycle. In the context of integrated, multi-domain, and multi-technology experimental networks, the focus of the SFI2 reference architecture, autonomy, and efficiency are critical requirements to orchestrate a complex and dynamic multi-domain virtual network composed of slices. The SFI2 reference architecture uses an ML-native approach to improve performance on complex decision-making problems throughout the slicing life cycle.

A network slicing system provides services with multiple security requirements. Providing various services in a multi-domain infrastructure for multiple customers makes security services crucial and complex. SFI2 leverages recent discussions, such as \cite{NGMN-Sec-Recommendations:16}, \cite{security:fatima22}, and \cite{3gpp_3rd_2021_ts28530}, to comprehensively address security issues for the slicing life cycle, i.e., preparation, commissioning, operation, and decommissioning, as well as for intra-slice and inter-slice communication.

So, the main contributions of the architecture proposed involve the enhancement of the following aspects:

\begin{itemize}
    \item \textbf{Artificial Intelligence and Machine Learning}. The proposed architecture uses machine learning techniques natively to improve network slicing in its phases, supporting the orchestration of network slices and prediction of resources and quality of service.
    
    \item \textbf{Energy Efficiency and Sustainability}. By focusing on strategies for resource allocation, SFI2 targets sustainability and energy efficiency at different stages of the slice life cycle (i.e., preparation, commissioning, and operation) according to their demands.
    
    \item \textbf{Security}. We clearly define security in the context and scope of slicing architectures since current frameworks can operate over complex domains involving diverse computing resources over a shared infrastructure. In this sense, SFI aims to assure isolation and integrity.
\end{itemize}

This article is structured as follows. Section \ref{sec:CurrentArch} presents network slicing architecture and project approaches. Section \ref{sec:SFI2Arch} details the SFI2 architecture functionality and building blocks. Section \ref{SFI2Enhancements} presents the SFI2 architectural enhancements regarding machine learning support, sustainability, and security. Section \ref{sec:SFI2Instantiation} focuses on instantiating the SFI2 architecture on the FIBRE experimental network, and Section \ref{sec:experimental_setup} presents an experimental scenario demonstrating the sustainability architectural enhancement. Finally, Section \ref{sec:conclusion} concludes with an overview of SFI2 main highlights, contributions, and future work.

\section{Current Network Slicing Approaches} \label{sec:CurrentArch}
\label{sec:preparation-phase}

Network slicing is an essential technological enabler for \ac{5GnB}, vehicle and drone communications, IoT deployments,  Telco clouds, Industry 4.0, Augmented Reality/Virtual Reality (AR/VR) technologies, and \ac{MEC}, among others. Some commonplace service characteristics in these areas include dealing with diversified and stringent user requirements, supporting dynamic and on-demand resource allocation, and the need for efficient resource orchestration and allocation \cite{wu_survey_2022}.

In this sense, network slicing provides a new virtualization approach for the components of wired and wireless networks, such as communication resources, \ac{RAN} connection, switching equipment, cloud infrastructures, and computation and storage resources. Network slicing allows the partitioning of physical and virtual resources with the capability to create, orchestrate, configure, and redefine the slice partitions as needed. Network slicing adds value to the networking perspective by allowing abstraction, isolation, orchestration, and softwarization of service deployments on network architectures \cite{ordonez-rollout:21}.

Network slicing is gaining a substantial research track with the proposals of network slicing-oriented architectures, standardization efforts, network slicing frameworks, and efficient network slicing solutions, typically based on machine learning techniques \cite{shen-ai-assisted-ns:20} \cite{9382026}.
Standardization institutions like the Third Generation Partnership Project (3GPP), European Telecommunications Standards Institute (ETSI), Internet Engineering Task Force (IETF), International Telecommunication Union — Telecommunication Standardization Sector (ITU-T), Open Network Forum (ONF), and Next Generation Mobile Network (NGMN) alliance are developing NS-oriented architectures. The network slicing architectures and standards proposed by these organizations are, to some extent, focused on or siloed in their actuation domains (telecommunications, 5G\&B, industry, Internet, others) and tend to anchor the service aspects prevalent in these domains, presenting divergences on their approaches \cite{necos_project_d31_2018}. In the following subsection, we describe the main initiatives of standardization institutions for network slicing.

\subsection{Network Slicing Standardization Initiatives}

The 3GPP network slicing architecture mainly addresses the 5G/6G wireless domain \cite{3gpp_3rd_2021_ts28530}. The architecture reflects that 3GPP is the main standard body for mobile communication networks, focusing on the next generation of mobile networks. The 3GPP architecture includes the slice concept and definition in terms of services chains with virtualized or physical resources, network components like access network, transport, and core, the end-to-end concept, and the life cycle of a network slice with four main phases: Preparation, Commissioning, Operation, and Decommissioning \cite{debbabi_5g_2021}.

The ETSI network slicing effort focuses on a general end-to-end next-generation network slicing (NGNS) framework and architecture for service providers. The intent is to coordinate and operate services as active network slices. The ETSI slice design includes a service-oriented approach, defining a slice abstraction, its reusability, and autonomy \cite{etsi_next_2018}. Significant results from ETSI efforts are Zero-Touch network and service management for management automation  \cite{benzaid_ai-driven_2020}.

The IETF NS architecture proposal accommodates network slicing definitions, services, components, and features in the IETF networking set of protocol, nomenclature, and recommendations ecosystem. In summary, the IETF NS architectural proposal defines the functionalities of deploying services by slicing physical and virtual resources and maps them to the overall and general IETF concept of management, control, and data planes. The NS IETF architectural focus is on the transport network part of the end-to-end network slices that involves, in addition, the edge (RAN) and core slices \cite{chahbar_comprehensive_2021}.

The Open Networking Foundation (ONF) network slicing architecture is based on the Software-Defined Networking (SDN) paradigm. In the ONF NS architecture proposal, the SDN controller and SDN clients are the key components. A slice is comparable to an SDN client isolated by the controller’s virtualization and client policy functions. The orchestration and the defined global policy functions control and optimize the slice. The overall slice controller is an SDN controller application \cite{open_networking_foundation_applying_2016}.

The ITU-T NS conceptual architecture consists of Logically Isolated Network Partitions (LINPs) over physical resources supporting network virtualization. LINP is an essential conceptual ITU-T architectural component. According to ITU-T Y.3011 \cite{itu-t_framework_2012}, slicing allows logically isolated network partitions with a slice considered a unit of programmable resources such as network, computation, and storage. A LINP is an isolated and programmable entity that provides users and service providers capabilities similar to traditional networks without network slicing.

The NGMN network alliance provides a network slicing conceptual architecture outline with three layers. It defines an end-user service or a business Service Instance Layer (SIL), a Network Slice Instance (NSI) providing a set of network functions required by a service instance, and a resource layer \cite{napolitano_network_2018}.

5G enhanced wireless mobile capabilities with massive Machine-Type Communication (mMTC), enhanced Mobile Broadband (eMBB), and Ultra-Reliable and Low-Latency Communication (URLLC) services are concentrating the research and development efforts around the world. Slicing is a fundamental tool for \ac{5GnB}; consequently, the most relevant projects concerning network slicing address the 5G scenario. Relevant research projects proposing NS approaches for 5G include the projects 5GEx, 5G‐SONATA, 5G‐PAGODA, 5G‐SLICENET, NECOS (Novel Enablers in Cloud Slicing), and 5Growth \cite{9422346}. The following subsection describes the main research projects for network slicing.

\begin{table*}[]
\caption{Network Slicing Characteristics for Standardization Initiatives and Network Slicing Projects.}
\label{tab:group1_v2}
\resizebox{\textwidth}{!}{%
\begin{tabular}{|c|c|c|c|c|c|c|c|c|c|c|
>{\columncolor[HTML]{F2F2F2}}c |}
\hline
\backslashbox[68mm]{Characteristics}{Initiative} & \multicolumn{1}{c|}{\textbf{3GPP}} & \multicolumn{1}{c|}{\textbf{ITU-T}} & \multicolumn{1}{c|}{\textbf{NGMN}} & \multicolumn{1}{c|}{\textbf{IETF}} & \multicolumn{1}{c|}{\textbf{ETSI}} & \multicolumn{1}{c|}{\textbf{ONF}} & \multicolumn{1}{c|}{\textbf{5G-SONATA}} & \multicolumn{1}{c|}{\textbf{5G-PAGODA}} & \multicolumn{1}{c|}{\textbf{5GEx}} & \multicolumn{1}{c|}{\textbf{NECOS}} & \multicolumn{1}{c|}{\cellcolor[HTML]{F2F2F2}\textbf{SFI2}} \\ \hline
\textbf{Multi-domain Slicing}                           & \faCircleO                          & \faCircleO                           & \faCircleO                          & \faCircle                           & \faCircleO                          & \faCircleO                         & \faCircleO                               & \faCircleO                               & \faCircle                           & \faCircle                            & \faCircle                                                   \\ \hline
\textbf{Slicing across Experimental Networks}           & \faCircleO                          & \faCircleO                           & \faCircleO                          & \faCircleO                          & \faCircleO                          & \faCircleO                         & \faCircleO                               & \faCircleO                               & \faCircleO                          & \faCircleO                           & \faCircle                                                   \\ \hline
\textbf{AI-Native Architecture}                         & \faCircleO                          & \faCircleO                           & \faCircleO                          & \faCircleO                          & \faCircleO                          & \faCircleO                         & \faCircleO                               & \faCircleO                               & \faCircleO                          & \faCircleO                           & \faCircle                                                   \\ \hline
\textbf{Security with Slicing-aware Capabilities}       & \faCircleO                          & \faCircleO                           & \faCircleO                          & \faCircleO                          & \faCircleO                          & \faCircleO                         & \faCircleO                               & \faCircleO                               & \faCircleO                          & \faCircleO                           & \faCircle                                                   \\ \hline
\textbf{Sustainability-aware Slice Building Capability} & \faCircleO                          & \faCircleO                           & \faCircleO                          & \faCircleO                          & \faCircleO                          & \faCircleO                         & \faCircleO                               & \faCircleO                               & \faCircleO                          & \faCircleO                           & \faCircle                                                   \\ \hline
\textbf{Marketplace for Slices}                         & \faCircleO                          & \faCircleO                           & \faCircleO                          & \faCircleO                          & \faCircleO                          & \faCircleO                         & \faCircleO                               & \faCircleO                               & \faCircleO                          & \faCircle                            & \faCircle                                                   \\ \hline
\textbf{Domain Customized Infrastructure Interfaces}    & \faCircleO                          & \faCircleO                           & \faCircleO                          & \faCircleO                          & \faCircleO                          & \faCircleO                         & \faCircleO                               & \faCircleO                               & \faCircleO                          & \faCircle                            & \faCircle                                                   \\ \hline
\textbf{Domain Customized Monitoring Interfaces}        & \faCircleO                          & \faCircleO                           & \faCircleO                          & \faCircleO                          & \faCircleO                          & \faCircleO                         & \faCircleO                               & \faCircleO                               & \faCircleO                          & \faCircle                            & \faCircle                                                   \\ \hline
\textbf{End-to-End Slicing}                             & \faCircle                           & \faCircleO                           & \faCircleO                          & \faCircleO                          & \faCircle                           & \faCircleO                         & \faCircleO                               & \faCircleO                               & \faCircleO                          & \faCircle                            & \faCircle                                                   \\ \hline
\textbf{Slice-as-a-Service}                             & \faCircle                           & \faCircleO                           & \faCircleO                          & \faCircleO                          & \faCircleO                          & \faCircleO                         & \faCircleO                               & \faCircleO                               & \faCircle                           & \faCircle                            & \faCircle                                                   \\ \hline
\textbf{Infrastructure Monitoring}                      & \faCircleO                          & \faCircle                            & \faCircleO                          & \faCircleO                          & \faCircle                           & \faCircleO                         & \faCircle                                & \faCircleO                               & \faCircle                           & \faCircle                            & \faCircle                                                   \\ \hline
\end{tabular}%
}
\end{table*}

In order to highlight our contributions, we summarize the related works in Table~\ref{tab:group1_v2}, where we represent relevant characteristics for the realization of network slices for standardization initiatives and slicing projects \cite{necos_project_d31_2018}. For standardization bodies, it is indicated only the explicit network slicing characteristic support. For this, we use the marker (\faCircle) to represent the achievement of the characteristic by the network slicing approach and the marker (\faCircleO) to represent the non-compliance of the characteristic.

\subsection{Network Slicing Projects}

The 5GEx project mainly aims to enable cross-domain orchestration of services over multiple or multi-domain single administrations \cite{bernardos_5g_2015}. The project focuses on designing a networking factory where a new network infrastructure and associated services are instantiated and deployed by software. In 5GEx, a slice manager supports the resource orchestrator functionality with multi-tenancy resources and multi-vendor \ac{VNF} management. 5GEx allows end-to-end network and service elements to mix in multi-vendor, heterogeneous technology, and resource environments.

The 5G SONATA project focuses on the flexible programmability of networks and the optimization of their deployments. A SONATA slice is a basic unit of programmability with a set of resources used in an end-to-end network service comprised of \acp{VNF}. The SONATA framework leverages the virtualization offered by cloud infrastructures and network programmability to develop and deploy network services and \acp{VNF} \cite{xilouris_devops_2017}.

The 5G-PAGODA project aims to create a scalable 5G slicing architecture by extending the current \ac{NFV} architecture to support network slices composed of multi-vendor \acp{VNF}. PAGODA architecture approach enables network flexibility and programmability to create and manage virtual network slices tailored to the needs of 5G verticals for \acp{MVNO} \cite{afolabi_towards_2017}. 
As a complementary objective, the 5G‐SLICENET project proposes an end-to-end intelligent network slicing approach with a slice management framework supporting virtualized multi-domain and multi-tenant in SDN/NFV-enabled 5G networks \cite{escolar_scalable_2020}.

The NECOS project creates a reference architecture for cloud-networking slicing enforcing the realization of the \ac{SlaaS} paradigm \cite{9422346}. The NECOS project addresses cloud and network slicing and considers the end-to-end and multi-administrative domain scenarios while embracing the \ac{SlaaS} paradigm. Beyond those essential characteristics, NECOS has unique features such as a new Virtual Infrastructure Manager (VIM) on-demand, a new Wide-area Infrastructure Manager (WIM) on-demand slicing models, and a marketplace approach, which other current approaches to slicing architectures have not considered.

Lately, projects such as 5G-Solutions~\cite{briguglio2021}, 5G-Tours~\cite{vignaroli2020}, 5G-Victori~\cite{mesogiti2021}, and 5Growth~\cite{li-xi-2021} have been proposed to validate the network slicing concept across multiple vertical industries, including transportation, energy, media, entertainment, and factories of the future. Among such projects, 5Growth has used artificial intelligence and machine learning solutions to enhance service automation in network slices. Also, Moreira \emph{et} al.~\cite{Moreira2021} designed and evaluated the NASOR, an orchestration concept towards multi-domain network slicing on top of Internet routers.

Current network slicing solutions envision highly relevant aspects: optimized resource allocation and orchestration, end-to-end network slicing, multi-technology, multi-tenant slicing support, automated slicing management, and elastic and dynamic resource allocation \cite{zhang_overview_2019}.
An open research issue currently addressed by network slicing architectures and projects such as 5Growth is an intelligent end-to-end slice orchestration and management solution that could fulfill diverse service requirements while ensuring efficient resource utilization and appropriate slice isolation.

The architectural enhancements proposed by the SFI2 architecture discussed in the following sections aim to add machine learning and security new capabilities to current network slicing architectures preserving a sustainable solution concomitantly.

\section{The SFI2 Network Slicing Architecture}
\label{sec:SFI2Arch}

\begin{figure*}[ht]
    \centering
    \includegraphics[width=.85\textwidth]{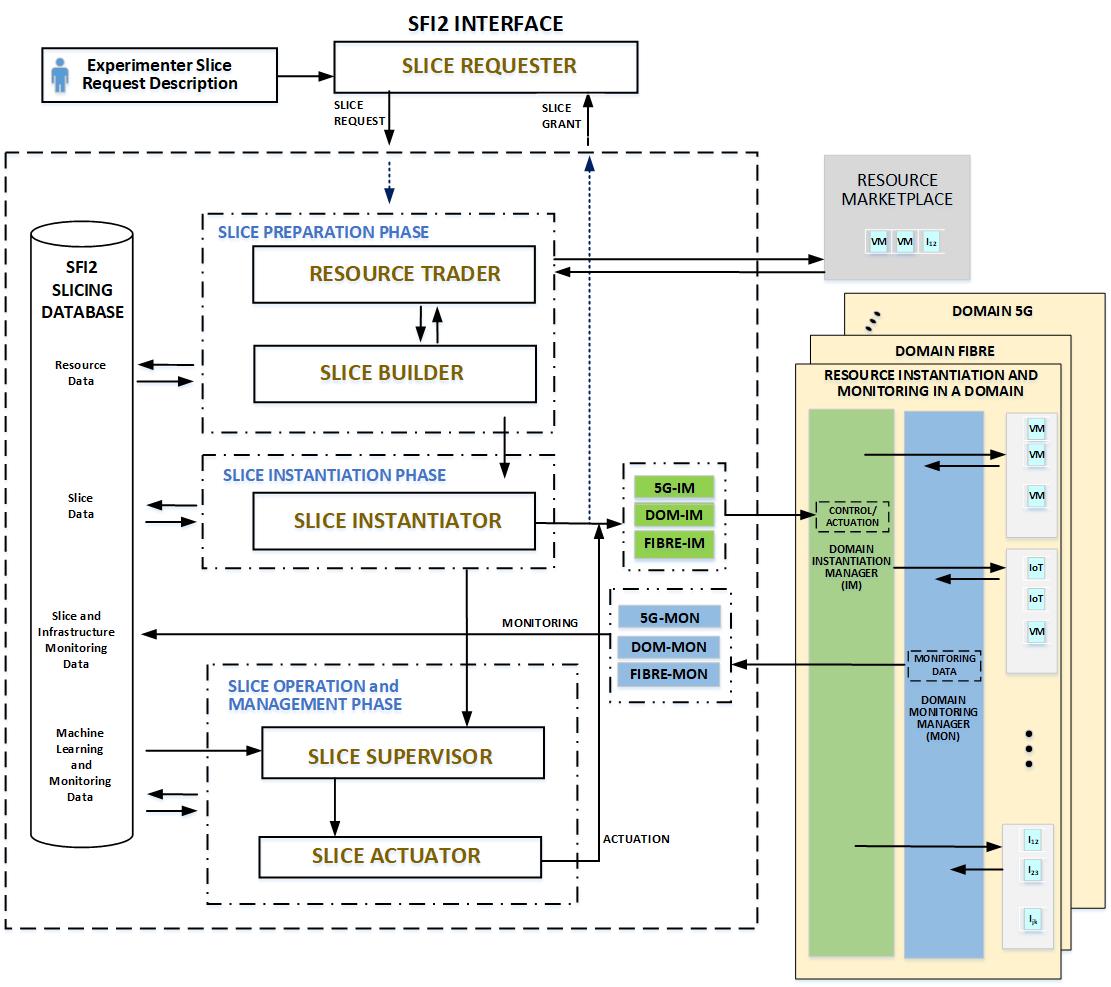}
    \caption{SFI2 Network Slicing Architecture and Functional Blocks.}
    \label{fig:SFI2SlicingArch}
\end{figure*}

The SFI2 network slicing architecture proposal is fundamentally a reference architecture that allows slicing functionalities for instantiation and deployment in distinct experimental network domains. The SFI2 architecture advances state-of-the-art by offering slice-as-a-service instantiation and deployment capabilities for multi-domain experimental networks.
In this context, the SFI2 architecture aims to fulfill an existing gap in providing a virtual experimental network using the network slicing paradigm across multiple experimental network domains. This integration aims, from the user's point of view, to support multi-technology experimental networks by integrating existing experimental networks that focus, in most cases, on specific technologies.

The SFI2 architecture offers dynamic slice building, instantiation, and supervision with machine learning-based embedded optimization. The SFI2 slicing capabilities have intradomain, interdomain, intraslice, interslice, and life cycle slicing security and an entirely new sustainable-aware approach for slice resource selection, orchestration, and deployment. SFI2 architecture's basic functionalities and modules are illustrated in Fig. \ref{fig:SFI2SlicingArch}. The architecture comprises a set of functional module components interrelated according to the slicing life cycle phases of preparation, commissioning, operation, and decommissioning \cite{3gpp_3rd_2021_ts28530}.

SFI2 has an external interface that allows experimenters and tenants to request the creation of sliced virtual networks. To make the facility's utilization easier, SFI2 adopts intent-based and script descriptions for network setup across single and multiple domains.
The resource marketplace plays an essential role in the network slicing process by, firstly, collecting resources from different providers in eventually different network domains and, subsequently, displaying and trading these resources to compose a sliced virtual network.

The SFI2 resource marketplace innovates compared to other existing resource marketplaces by aggregating and trading multi-technology resources like virtual machines, SDN-capable switches, IoT resources, and 5G setups, among others, to compose a single virtual network. These resources are currently available on specialized and independent experimental networks, and SFI2 groups them in a structured way. At this point, it is important to remark that the SFI2 deployment focus is the set of experimental networks currently available in Brazil: FIBRE-NG \cite{abelem_fit_2013}, CloudNEXT \cite{atmosphere_cloudnext_2019}, FUTEBOL \cite{both_futebol_2019}, FIWARE \cite{oliveira_improving_2018}, 5GINFIRE \cite{silva_5ginfire_2019}, and NECOS \cite{9422346}.

The preparation of the sliced virtual network, composed of multi-technology resources belonging to eventually distinct domains, is the task of the resource trader and slice builder. The resource trader and slice builder modules cooperate to orchestrate the slice preparation by finding resources and allocating them to a specific sliced network.
In the SFI2 architecture, the building process, composed of slice resource trading and slice orchestration modules, considers user-defined sustainability and energy-efficiency parameters and constraints. On top of that, resource orchestration is optimized by using machine learning algorithms.

The slice instantiation module carries out the instantiation phase. It effectively deploys the assigned resources of a virtual network on their respective domains. The instantiation involves all domain-related aspects and specificities and includes all network slices and communication facilities required to allow slice operation.
The per-domain instantiation uses specific domain interface managers (DOM-IM) customized for each domain involved in the slicing process. From an architectural point-of-view, the SFI2 use of the DOM-IM modules approach allows an agnostic resource deployment in different domains.

The slice operation and management phase of the network slicing process is executed by the slice supervision, slice actuator, and monitoring interface managers (DOM-MON).
Slice supervision in the SFI2 architecture consists of dynamically verifying per-slice conformance of actual slice key performance indicators (KPI) concerning the user-defined slice key performance parameter (KPP) and management requirements like high-level SLA, QoE, or QoS specifications \cite{2022-ETSI-5G-KPI, 2019-KPI-5Gns}.

An innovative supervision capability of the SFI2 architecture is its ability to monitor application-specific performance parameters. As illustrated in Fig. \ref{fig:SFI2SlicingArch}, the architecture allows monitoring application-specific characteristics at the slice level. The application level monitoring facility (APP-MON) notifies the slice supervision module of alerts to allow on-the-fly slice reconfiguration.

Slice supervision in the SFI2 architecture uses machine learning techniques to support on-the-fly slice reconfiguration and slice elasticity. The slice actuator module executes the reconfiguration and elasticity required deployments at the request of the slice supervision module.

Monitoring is an essential part of any shared dynamic environment such as \ac{SFI2} in which several simultaneous users and tenants request network, processing, and memory resources and expect that specific quality, performance, security, and energy efficiency levels be met during the lifetime of their experiments and service deployments.
Monitoring in SFI2 architecture attempts to be agnostic. The performance parameters monitoring data can be generated and collected by several tools, such as Prometheus, Casandra, and others, at each infrastructure and resource provider domain. The monitoring interface managers (DOM-MON) collect the required monitoring data at the domain and transfer them to SFI2 modules processing providing the required customization of the acquired data.

\begin{figure*}[ht]
    \centering
    \includegraphics[width=.8\textwidth]{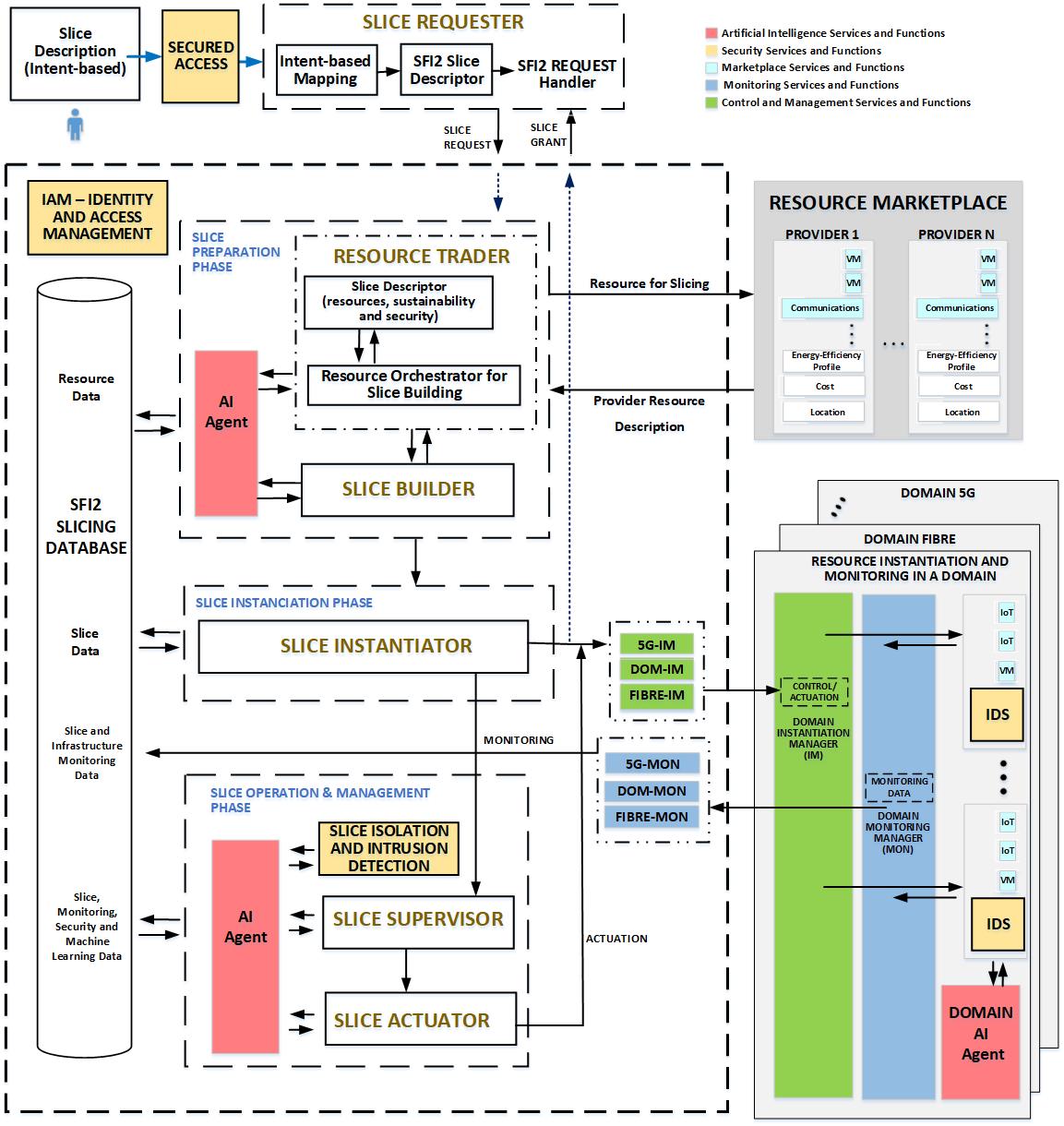}
    \caption{SFI2 Machine Learning, Sustainability, and Security Architectural Enhancements.}
    \label{fig:Enhancements}
\end{figure*}

The SFI2 slice database supports the entire slicing process in the preparation, instantiation, slice operation, and management phases. The SFI2 slicing database stores and shares monitoring data, slicing resources and slicing deployment information among the SFI2 modules. The SFI2 database also supports the security functionalities that encompass the set of activities concerning user access and the SFI2 module's operation.

The main contribution of the SFI2 network slicing architecture is to enhance current slicing architectures with new fundamental capabilities. The new capabilities enhancements are threefold:

\begin{itemize}
     \item Slice creation and operation with sustainable resource allocation;
     \item Resource trading, slice building, and slice operation elasticity with machine learning and optimization techniques support (AI-native architecture); and
    \item Security features and functionalities embedded within slice operation and experimental network domain deployment.
\end{itemize}

In summary, the SFI2 architecture aims to create slices for experimental networks, including operation and management optimization, while considering the utilization of sustainable resources in a secure environment, as discussed in the following sections.

\section{SFI2 Functional Enhancements} \label{SFI2Enhancements}

This section gives details of the architectural enhancements of the SFI2 architecture.

\subsection{Artificial Intelligence and Machine Learning Enhancements}

The SFI2 architectural machine learning enhancements are illustrated in Fig. \ref{fig:Enhancements}. In the life cycle of network slices, some phases are foreseen, such as preparation, commissioning, operation, and decommissioning. The SFI2 proposal is to use machine learning techniques natively in the architecture to improve network slicing in some of these phases \cite{Wu2022}.

Among the enhancements supported by machine learning techniques in the orchestration of network slices, resource prediction and quality of service prediction stand out. Different techniques ranged from combinatorial optimization to machine learning, such as reinforcement learning and supervised and unsupervised learning. Notably, their applications predominantly focused on commissioning, instantiation and monitoring, and control for all the techniques considered. However, intelligent and autonomous actions based on planning on already deployed network slices lack innovation, leading SFI2 to propose an intelligently-native orchestrator for the entire network slicing life cycle.

Furthermore, the SFI2 orchestrator aims to operate over heterogeneous infrastructures to support native distributed machine learning on its building blocks \cite{Nasir2021, Domeke2022}.
SFI2 will follow the concept of \ac{MLaaS} using distributed agents. To do so, SFI2 builds ML-based management that interacts with the monitoring system and with other blocks of the network slicing management. The ML management module maintains already-trained machine learning models and carries out distributed training between target domains for building specific models and up-to-date ones.

The SFI2 ML manages learning agents throughout all infrastructure components to support distributed machine learning. Those agents can perform training or prediction tasks coordinated by the principal ML agent. In our orchestration framework, we are pursuing a mechanism capable of enclosing numerous machine-learning paradigms and using them depending on the network slicing life cycle phase.

\subsection{Energy Efficiency and Sustainability Enhancements}

Adopting energy-efficient and renewable-energy-aware practices is a prime goal for any organization in the \ac{ICT} industry concerned with global sustainability.
The amount of CO2 generated by \acp{ICT}, which  includes data centers, networks, and end-user equipment, has been assessed.
According to GeSI --- Global Enabling Sustainability Initiative, the contribution of \ac{ICT} is about 1.25 Mton of CO2 per year~\cite{strategy2015smarter2030}. 
Despite the increasing number of computing devices, this number has been gradually decreasing due to actions regarding the energy efficiency of the equipment and its operation. 
In this sense, the deployment of network slices optimizes computing and network resources through their virtualization. 
However, it requires more complex functions and management. 
Therefore, energy efficiency has to be targeted at different stages of the slice life cycle, including slice preparation, commissioning, and operation, which have varying energy demands. 
Beyond that, tenants in the \ac{ICT} business may explicitly require specific sustainability (\textit{e.g.}, clean/renewable energy sources) and energy efficiency indicators due to their commitments to users and investors adherent to the \ac{ESG} principles~\cite{bassen2020environmental}.

The SFI2 architecture works toward a sustainable and energy-efficient network slicing process, addressing features aligned with cloud computing and general data centers' equivalent policies.
As a general policy, service providers' resource allocation, including slice deployments, should always favor the least energy consumption, ideally coming 100\% from renewable sources~\cite{2019-Khattar}.

The design of the SFI2 architecture addresses a perspective that energy (in terms of  electricity power) is a resource as essential and scarce as computational resources.
For instance, this perspective allows one to approach energy usage the same way as CPU usage.
Therefore, SFI2 adopts resource allocation strategies considering the energy usage of the final system, aiming to reduce it as much as possible. 
While also ensuring that slice requirements are attended to.
Moreover, the energy-as-a-resource perspective enables SFI2 to follow policies and guidances considering the energy sources (\textit{e.g.}, renewable, non-renewable).

The SFI2 architecture endorses state-of-the-art policies and methods to achieve energy efficiency for distributed systems, focusing on strategies for resource allocation.
In this direction, we list the following strategies considered in the SFI2 architecture:
\begin{itemize}
    \item Building network routes/paths aiming to reduce the communication cost;
    \item Enforcing efficient computational resource allocation;
    \item Applying local solutions to increase energy efficiency (\textit{e.g.}, switches with appropriate technology, sleep-mode mechanisms); and
    \item Prioritizing allocating resources from providers using renewable energy.
\end{itemize}

From a conceptual-architectural perspective, SFI2 adopts such strategies by improving some of the fundamental slice life cycle components, namely, resource marketplace and slice builder (Fig. \ref{fig:Enhancements}).
The resource marketplace is in charge of grouping all the available resources required to build a virtual network by sliced resources, indicating the available resources and their description (\textit{e.g.}, type of resource, resource owner, price per hour, flag for renewable energy usage, available amount).
In this sense, the resource marketplace serves as a menu for computational resource selection fitting the tenant's requirements.

The slice builder module uses machine learning to analyze the resource list available in the marketplace, looking for an optimal decision on the resource selection for the slice deployment on domains towards a virtual network deployment.
A rating scheme favors the resource providers showing better \ac{PUE}, greater renewable consciousness, and the lowest communication cost (in terms of geographical distance and number of hops). 
The slice instantiation module is then responsible for performing the actual deployment task.

\subsection{Security Enhancements}

\begin{table*}[ht]
\centering
\caption{Attack Categories and Network Slicing Phases.}
\label{tab:attack-cat}
\begin{tabular}{|c|c|c|c|c|}
\hline

\multirow{1}{*}{\textbf{Attack Categories}} & \textbf{Preparation} & \textbf{Commissioning} & \textbf{Operation} & \textbf{Decommissioning} \\ \cline{2-5} 
                            \hline
Impersonation & \faCircle & \faCircle & \faCircle & \faCircle  \\ \hline
Traffic Injection & \faCircle & \faCircle & \faCircle & \faCircle  \\ \hline
Denial-of-Service (DoS) & \faCircle & \faCircleO  & \faCircle & \faCircleO  \\ \hline
Tampering & \faCircle & \faCircle & \faCircle & \faCircle  \\ \hline
Eavesdropping & \faCircle & \faCircle & \faCircle & \faCircleO  \\ \hline
Reply Attacks& \faCircle & \faCircleO & \faCircleO &  \faCircleO  \\ \hline
Interfaces Monitoring & \faCircle & \faCircle & \faCircle & \faCircle  \\ \hline

\end{tabular}
\end{table*}

Security is often overlooked or even neglected in network testbed research, which primarily focuses on its physical and functional aspects. The reason for that is perhaps the same that led the Internet architecture to bring on security at a later phase. Testbed architects' academic mindset usually assumes the legitimate use of its resources and focuses on delivering facilities rather than preventing or protecting them against misuse. However, for 5G security, several prominent bodies, such as NGMN, are concerned with this issue and are making recommendations to be considered at the design stage of network slicing \cite{NGMN-Sec-Recommendations:16}. Conversely, the SFI2 security challenge is how to handle the integration of testbed architectures originally built with only minor security capabilities.

The proposal for SFI2 architecture with security enhancements is illustrated in Fig. \ref{fig:Enhancements}. Firstly, we must present a clear definition of security in the context and scope of the SFI2 architecture, since the term can be taken by its diverse meanings. Slicing frameworks operate over complex domains of resources, each of which bears its vulnerabilities and special features regarding security. In addition, the more performance decoupling among slices sharing a given physical infrastructure is promoted, the better is a network slicing solution \cite{slice-isolation:18}. This implies that multiple slices over a shared infrastructure can coexist without much interference among them, assuring isolation, and, thus integrity.

So, through a synthetic definition, we define security via five primary requirements: Confidentiality, Integrity, Availability, Authentication, and Authorization \cite{net-slicing-security:19} \cite{Security:Wang:2009}. And these requirements should be considered along different phases of a slice life cycle. 

As stated earlier, the slicing life cycle is composed of 4 main phases: preparation, commissioning, operation, and decommissioning \cite{3gpp_3rd_2021_ts28530}. The literature identifies inherent vulnerabilities in each phase \cite{security:fatima22}. Among the main contributions of the SFI2 architecture, one is to consider such vulnerabilities and propose architecturally-embedded solutions against multiple attack vectors throughout the slice life cycle. 

Table~\ref{tab:attack-cat} presents the slice life cycle phases and attack categories that can impact them. We use the marker  \faCircle $ $ to denote the attack accomplishment on the slicing life cycle; otherwise, we use \faCircleO $ $  where the attack class does not impact the slicing life cycle. Different attack classes may target distinct life cycle phases, as follows:
\begin{itemize}
    \item Impersonation: Attack that attempts to bypass authentication and authorization procedures. It is an attempt by an unauthenticated and/or unauthorized user to make a request to the system and to be attended by it;
    \item Traffic Injection: It is the insertion of foreign - unusual or invalid - traffic to the components of the architecture aiming at  generating greater consumption of bandwidth on the network and increased processing on servers and switches. This class of attack is related to the Denial-of-Service class, and it is mostly intended to compromise availability;
    \item Denial-of-Service (DoS): This is an attempt to make a service unavailable to legitimate users, temporarily or indefinitely. It can occur in a variety of forms, the most prominent being the flooding of resources with superfluous requests and injection  of foreign traffic;
    \item Tampering: This is the unauthorized interference over code, data, or physical device  of a system intended to modify or manipulate it. Integrity is the first requirement affected by this attack, which can even lead to service unavailability;
    \item Eavesdropping:    This is the attack that attempts to access information in the communications between communicating entities by capturing data. Such kind of attack impacts the confidentiality and even the integrity of the targets. 
    \item Reply Attacks: This type of attack aims to capture access control information previously sent from one entity to another and its subsequent retransmission with the intention of producing an unauthorized effect or gaining unauthorized access. In this way, the goal of the attack is to perform operations originating from legitimate users, targeting, for example, to affect the availability and integrity requirements;
    \item Interfaces Monitoring: This attack targets the interfaces of a system, e.g.: northbound and southbound, aiming to control the  interfacing components. It impacts the confidentiality requirement at first since it captures any and all information. After that, other malicious activities  can be performed.
\end{itemize}

The impersonation attack (Table \ref{tab:attack-cat}) can be performed at all stages of a slice's life cycle. To prevent this attack, the SFI2 architecture provides, as presented in Fig. \ref{fig:Enhancements}, an element called IAM (Identity and Access Management). This element is responsible for the management of the identity of the involved parts as well as for the control of access to services and resources. The IAM performs identity authentication  and access authorization in the SFI2 architecture, for the experimenter/tenant of the architecture as well as for all elements that compose SFI2. It aims to ensure that all elements that interact or provide service to SFI2 are authenticated and  operations are authorized.

The traffic injection attack (Table \ref{tab:attack-cat}) can also be performed at all stages of the life cycle. The SFI2 architecture mitigates the risk of suffering from such a class of attack by blocking any and all traffic coming from the network external to the architecture's network. In addition, all elements of the SFI2 architecture will only accept traffic generated by the architecture elements themselves. This blocking will be accomplished with firewall rules and also for the controls of the IAM.

The DoS attack, as shown in Table \ref{tab:attack-cat}, can be executed in the preparation and operation phases. During the preparation phase, the attack may intend to deny the tenant - which is also the experimenter - the ability to request the slice creation. Therefore, the API/Portal of the SFI2 architecture must be equipped with mechanisms that prevent or mitigate this attack. To this purpose, SFI2 uses IDS (Intrusion Detection System) and IPS (Intrusion Prevention System) mechanisms combined with packet filtering. In addition, this architecture element can be provided with high scalability and also with autoscale mechanisms \cite{security:correa19}. While in the operation phase, the goal of the DoS attack is to deny the slice service to the experimenters/tenants on the SFI2 architecture.  To prevent this attack, the SFI2 architecture uses IDS/IPS mechanisms and packet filtering on slice accesses.


The tampering attack can be performed in all phases of the slice life cycle. For this attack, the SFI2 architecture only changes information or code of any element by a token generated for this purpose by the IAM. Thus, the user/element that wants to make any change must request a token from the IAM, while the element that will receive the change must check with the IAM if the token presented is a valid token.

The eavesdropping attack can be performed in the preparation, commissioning, and operation phases. To avoid this attack, it is a requirement in the SFI2 architecture to use cryptography between the communication pairs. The exchange of the cryptographic key must be performed at the moment of authentication of the elements with the IAM. The use of cryptography by all elements of the architecture also aims to prevent the attack on monitoring interfaces, which can reach all phases of the slice life cycle.

In the SFI2 architecture, each new request from a user or an element of the architecture to another element must be made using a token. In this way, a new token will be generated for each new request. This creation and control of the tokens are executed by the IAM. This functionality aims to prevent reply attacks.

Additionally, the SFI2 slicing database component (Fig. \ref{fig:Enhancements}) is provided with mechanisms to ensure the integrity of the information stored in this database. For this, any insertion or removal of data from the database must be authorized by the IAM. In addition, the database itself must be equipped with redundancy to maintain the persistence of the slices' information.

Finally, as a security mechanism, the SFI2 architecture, as presented in Fig. \ref{fig:Enhancements}, has a specific database, called "SFI2 Security Database", in which all the logs of security-related incidents and activity are stored. This database is used for auditing and also as an input to machine learning algorithms, to learn about new attacks and to suggest mitigation methods.

The attacks and solutions discussed above are in the context of the architecture domain. Thus, in addition to attacks related to the slice life cycle, there are security issues internal to the slice and in communication between the multiple slices. Some of these security issues are mitigated by a single point of access to the architecture/slice. That is, the experimenter and end-user use the same environment for management, configuration, and access to the slice. In this way, the SFI2 architecture can have access control and insert security levels. It is assumed that an experiment can request end-to-end communication, including being a request on the slice template at creation time. With the single point of access, this request from the experimenter can be met with the integration of the endpoint device (the gateway) and how the end client connects to that gateway (e.g. VPN). So, as these issues are pertinent, they will be discussed soon, with the current SFI2 architecture being an experimental system, with security updates focused on the slice life cycle.

\begin{figure*}[ht]
    \centering
    \includegraphics[width=0.85\textwidth]{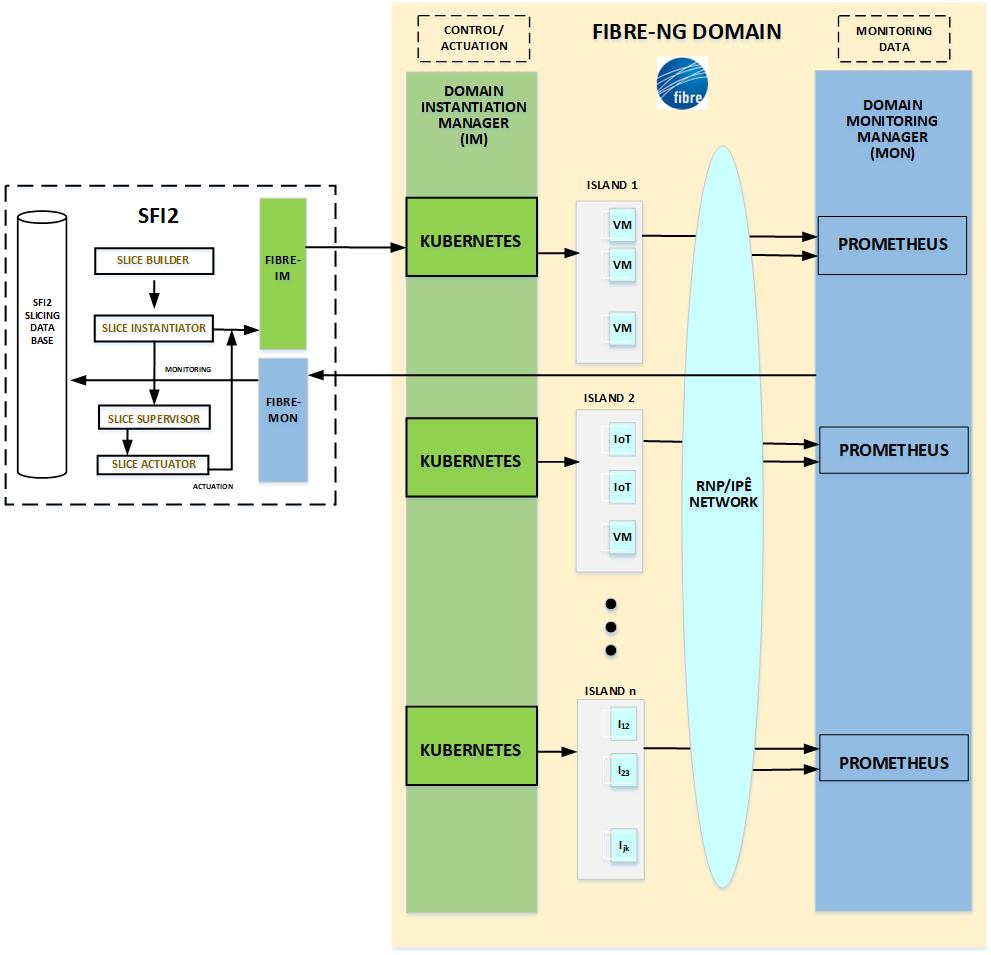}
     \caption{SFI2 Architecture Instantiation and Deployment in FIBRE-NG Domain.}
    \label{fig:SFI2-FIBRENG}
\end{figure*}

\section{SFI2 Instantiation and Deployment over Experimental Network Domains} \label{sec:SFI2Instantiation}

\begin{figure*}[ht]
    \centering
    \includegraphics[width=.99\textwidth]{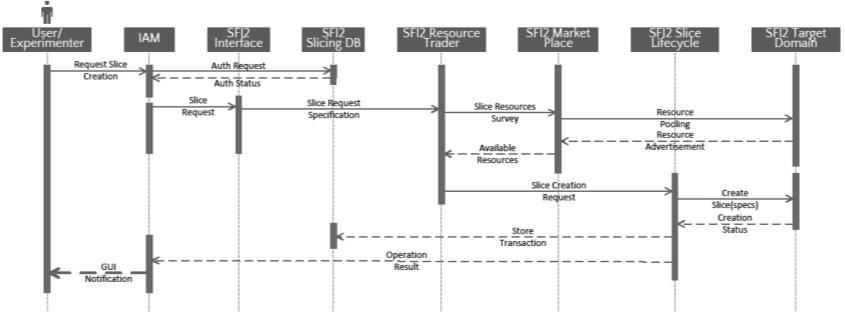}
     \caption{Slicing Creation Flow Interactions.}
    \label{fig:slice_creation_flow_sequence}
\end{figure*}

\color{black}

This section illustrates how SFI2 reference architecture is instantiated and deployed over a specific experimental network domain and highlights how the reference architecture addresses multi-domain and multi-technology aspects when integrating multiple experimental networks.

One of the contributions of the SFI2 project is its capability to integrate existing experimental testbeds scattered around different institutions (universities, research centers, non-governmental institutions, and others) into a single, multi-domain experimental testbed. Each integrated testbed will run an instance of its domain management components so that SFI2 experimenters can book resources like containers, virtual machines, bare metal, and IoT devices in these instances. The SFI2 orchestrator handles these resources to provide network slicing on top of SFI2 target domains.

Fig. \ref{fig:SFI2-FIBRENG} illustrates the physical and logical elements, modules, and components involved in the instantiation of the SFI2 reference architecture for the FIBRE-NG experimental domain \cite{abelem_fit_2013}. 

The FIBRE-NG domain is an experimental network composed of islands hosting resources (VMs, containers, and bare metal) for experimentation and the IPÊ network ~\cite{Zanotelli2021}. The islands are physically located in universities and research centers scattered in Brazil from north to south, and the IPÊ network provides the necessary network interconnection among islands to support running distributed experiments. FIBRE-NG has 18 active islands, allowing an experimenter to set up and run distributed experiments on a geographically extensive real network. Regarding technology, the FIBRE-NG supports container deployment based on Kubernetes \cite{raj_kubernetes_2023} and has a monitoring infrastructure based on Prometheus \cite{sukhija_towards_2019} for its geographically distributed islands and the IPÊ interconnection network.

At this point, it is essential to highlight that the FIBRE-NG experimental network has independent setup and experiment running support for experimenters \cite{e_nacional_de_pesquisa_fibre_2022}. As such, the SFI2 instantiation and deployment in the FIBRE-NG domain is an additional alternative to configuring resources for experimentation. SFI2 slicing over the FIBRE-NG will complement and add new resources and benefits to FIBRE-NG users by incorporating ML-native optimization, energy-efficient resource utilization, and slicing-tailored security functionalities. Beyond that, the integration with other experimental domains will allow experimenters to allocate services, platforms, 5G functions, and IoT device resources hosted by other experimental domains.

As illustrated in Figure \ref{fig:SFI2-FIBRENG}, the SFI2 deployment over the FIBRE-NG experimental network uses a domain interface manager (DOM-IM) based on Kubernetes (FIBRE-IM) to allocate and deallocate slicing resources such as VMs, containers, and bare-metal in the FIBRE-NG islands.

The SFI2 resource orchestrator (Figure \ref{fig:SFI2-FIBRENG}) enhances and takes precedence over the FIBRE-NG Kubernetes orchestration stream to handle the management and lifecycle of network slices. Accordingly, slice instantiation is achieved through the Kubernetes-aware FIBRE-IM, a DOM-IM module customized for this specific domain. The modular DOM-IM approach of the SFI2 architecture agnostically facilitates the integration of distinct experimental domains. Since SFI2 has ML-native optimization features, slice supervision may dynamically reconfigure slices in the FIBRE-NG domain using the FIBRE-IM or, in any other domain, making use of its specific DOM-IM. In general, the SFI2 resource orchestrator has specialized functions such as deployment in target domains with heterogeneous virtualization technology, full-stack monitoring requirements, and specific configurations for enabling inter-domain isolation.

The domain monitoring interface of SFI2 (DOM-MON) (Figure \ref{fig:SFI2-FIBRENG}) is the abstraction used to agnostically monitor the components involved in the creation, operation, and management of slices. In the context of the FIBRE-NG domain deployment, the FIBRE-MON module is the element that interfaces SFI2 modules with the Prometheus FIBRE-NG native monitoring system. FIBRE-MON module allows the monitoring of the islands, allocated virtual machines,  bare-metal,  and the interconnection backbone provided by the IPÊ network.

When integrated through the SFI2  reference architecture, other experimental testbeds will provide different resources like IoT and drones on the SFI2 marketplace. For example, an experimental testbed focusing on 5G may provide resources related to the 5G Core and the 5G Radio interface. In contrast, a cloud testbed may provide common platforms as a service or even machine learning such as Jupyter notebooks and compute-intensive applications such as Apache Spark and Apache Kafka.

The resource heterogeneity inherently existing in multi-domain and multi-technology experimental networks poses challenges over the current state-of-the-art slicing orchestration, especially because end-to-end network service has stringent requirements. The SFI2 architecture addresses these challenges by adopting natively intelligent orchestration, as illustrated in the discussed FIBRE-NG instantiation and deployment.

\section{Experimental Assessment}\label{sec:experimental_setup}

Aiming to validate our architectural enhancements, we propose a two-fold experimental scenario to assess the functionalities of our slicing orchestration architecture. This testing scenario showcases how AI techniques are essential for secure, sustainability-aware network slicing architectures.

Our experimental flow follows Fig.~\ref{fig:slice_creation_flow_sequence}, which depicts the deployment flow diagram of network slicing through the SFI2 Architecture. Initially, we outline the enhancements in security within slice orchestration architectures achieved through the utilization of native and distributed machine learning agents within the architectural blocks. Subsequently, we discuss the advancements in the context of sustainability.

\begin{figure*}[h]
    \centering
    \includegraphics[width=.9\textwidth]{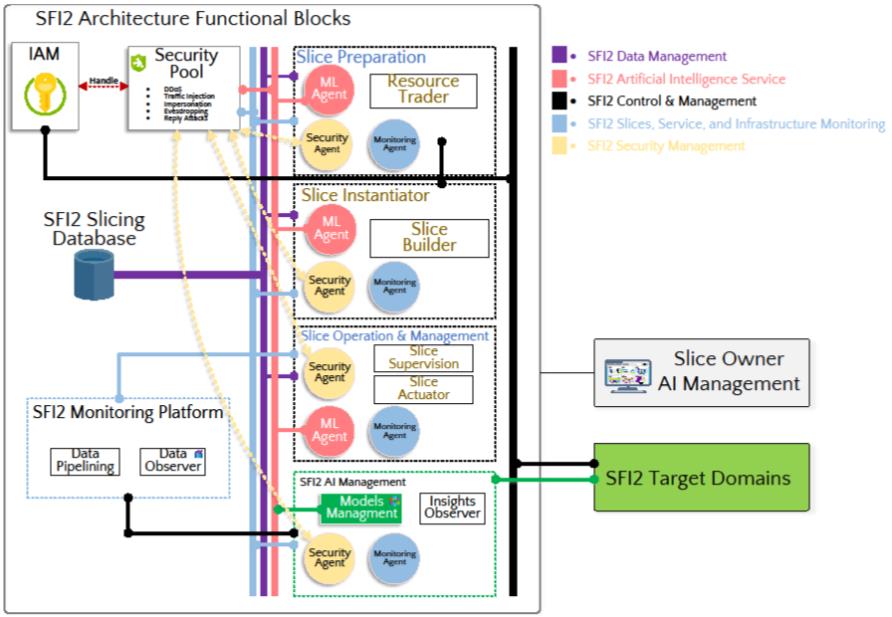}
     \caption{Security Enhancement for SFI2 Reference Architecture.}
    \label{fig:security_experimental_setup}
\end{figure*}

\subsection{Security}\label{subsec:security_enhancements}

We built an experimental scenario to evaluate the security enhancement of the SFI2 Architecture that combines \ac{ML} techniques with a distributed security approach oriented to the life cycle of the network slice. Fig.~\ref{fig:security_experimental_setup} represents with more detail the conceptual blocks of the SFI2 Architecture. It highlights the network slice orchestration and security blocks that are the \ac{IAM} and Security Pool. The SFI2 Architecture has different data, control, and management planes for the different services and architectural blocks.

The SFI2 Security Management plane is responsible for security control and enforcement in the functional blocks of the SFI2 Architecture. In this security control plane, asynchronous interaction exists between it and all the available blocks of the SFI2 Architecture. The \ac{IAM} Block is responsible for the authentication and authorization of users and the functional blocks that make up the SFI2 Architecture. \ac{IAM} interacts with the Security Pool, enabling the architecture manager or the owner of the network slice to use the necessary security mechanisms for its deployment.

Our architectural framework can deal with the following security threats: DDoS, traffic injection, impersonation, tampering, eavesdropping, reply Attacks, and interface monitoring. The current literature regarding slicing architectures does not cover these security aspects widely. We use a distributed security mechanism natively integrated with the functional blocks of the architecture. The interaction between the security mechanisms with the \ac{ML}-Agents distributed throughout the architecture and its functional blocks is responsible for this distributed approach. An \ac{ML}-Agent works alongside the slicing orchestration blocks enabling them to apply predictions, forecasting, and classification on demand, acting as an \ac{API} of \ac{ML} services for the architecture block. Thus, an \ac{ML}-Agent in Slice Preparation acts differently from the \ac{ML}-Agent present in the Slice Instantiation block.

The SFI2 AI Management Block is where the architecture manager and the network slice owner manage the AI services the architecture supports. Thus, the selection of an AI service for each \ac{ML}-Agent, as well as the training of the \ac{ML} algorithm, takes place in this functional block~\cite{Moreira2023}. Additionally, the architecture provides the SFI2 Monitoring Platform block, which monitors each component of the architecture and the orchestration services that run on top of it. Thus, to validate the distributed security mechanism that the SFI2 Architecture brings, we evaluated the empowering of \ac{DDoS} Defense in the architecture blocks. This security mechanism aims to empower the functional blocks of the architecture against DDoS attacks that can lead to its instability or unavailability.

\subsubsection{Experimental Scenario}\label{experimental_scenario_security}

\begin{figure*}[ht]
    \centering
    \includegraphics[width=.6\textwidth]{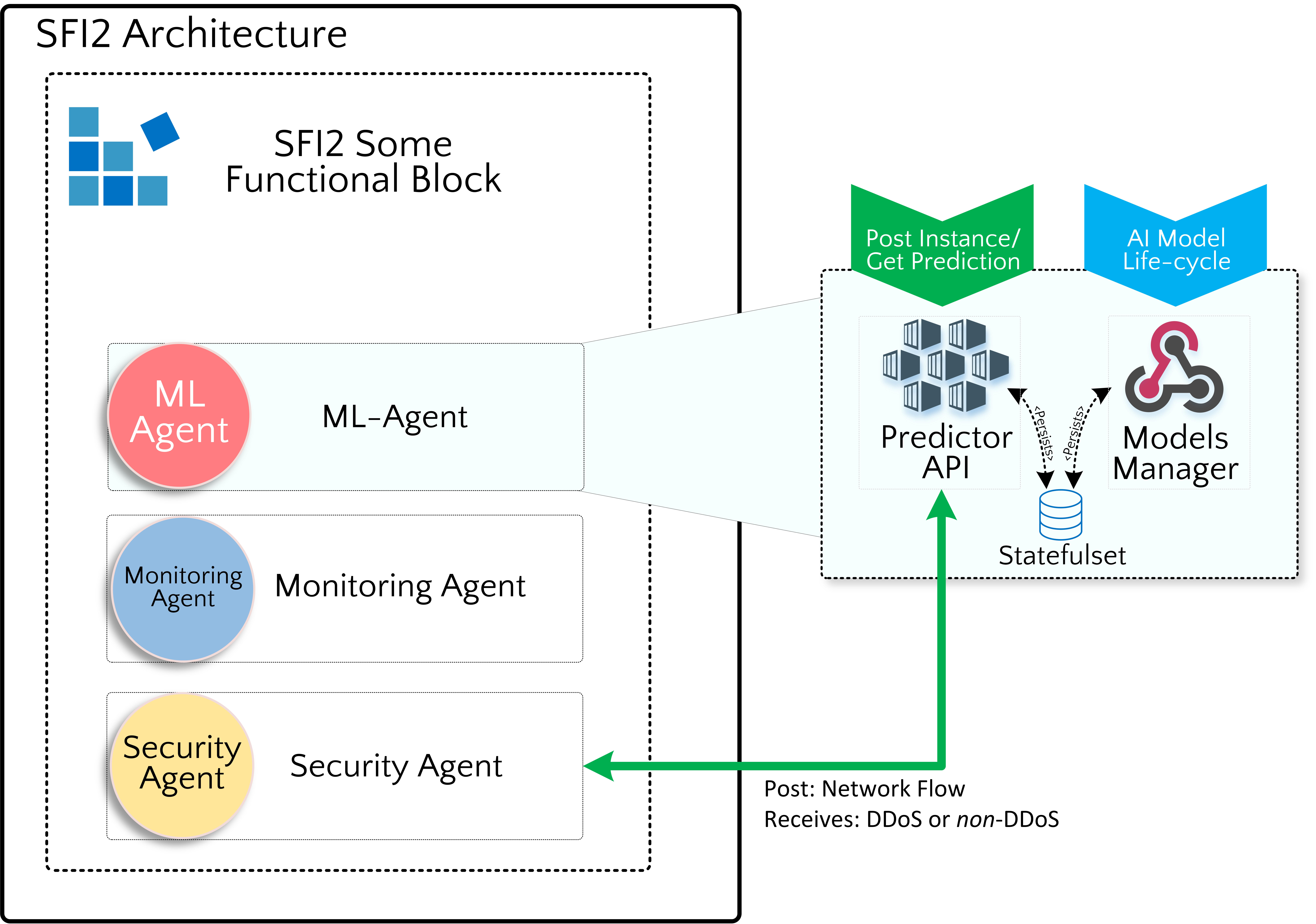}
     \caption{ML-Agent Specification and Functionalities.}
    \label{fig:security_experimental_scenario}
\end{figure*}

We propose the experimental scenario in Fig.~\ref{fig:security_experimental_scenario} to functionally validate our hypothesis of building a slicing architecture with native \ac{AI} and security. For this, as Fig.~\ref{fig:security_experimental_scenario}, we take a functional block of the architecture and exploit its roles and interfaces of the \ac{ML}-Agent. ML-Agent provides two types of services, one is provided by the Predicor \ac{API} which comprises of an \ac{API} which takes data as input and returns a \ac{ML} prediction as a response. The other functionality of the AI-Agent comes through the AI Model Life-Cycle, which is where the AI-Agent interacts with the SFI2 AI Management (Fig.~\ref{fig:security_experimental_setup}) to insert, remove or update models from \ac{ML} accordingly.

\begin{figure}[ht]
    \centering
    \includegraphics[width=.99\columnwidth]{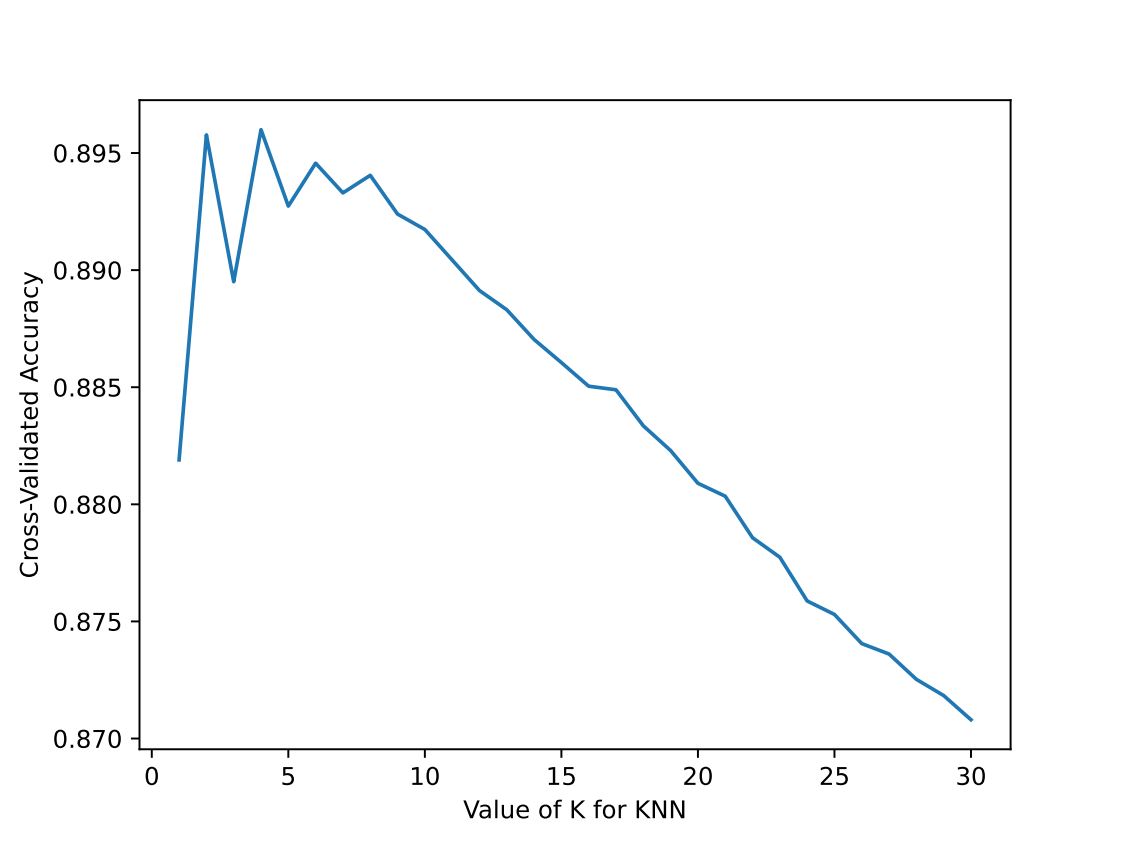}
     \caption{\ac{KNN} for \ac{DDoS} Attacks Prediction.}
    \label{fig:knn_cross_validation}
\end{figure}

In this experiment, using SFI2 AI Management, we trained some \ac{ML} models with a \ac{DDoS} dataset in order to finally have an accurate \ac{ML} model capable of responding to a given network flow if deals with \ac{DDoS} and or benign traffic. At the end of the training, the model was exported and loaded into the \ac{AI}-Agent through the ML-Agent's AI Model Life-Cycle interface. At this point, the Security Agent can query the Predictor \ac{API}, providing a sample of the network flow and receiving as a response the classification of the network flow according to whether the profile is \ac{DDoS} or benign.

\subsubsection{Dataset}\label{subsubsec:security_dataset}

For that, we trained several \ac{ML} models using the AI Management block considering the DDoS-2019~\cite{Sharafaldin2019} dataset. This dataset contains packets representing DDoS attacks in raw \textit{pcap} format. Thus, using the FlowMeter~\cite{Arash2017}, a flow sampling tool, which is a time-based features extractor tool, we train our \ac{ML} algorithms on these collected flows. The flows that the tool extracted contain more than 80 time-based features with many \ac{DDoS} attacks, making it possible to apply \ac{ML} algorithms to extract patterns from these features. Furthermore, once we train the models through SFI2 AI Management, it is possible to export the trained model to the ML-Agents that are distributed throughout the architecture to carry out predictions, forecasting, and classifications for the functional blocks they require.

Each architectural block contains a mechanism for sampling packets and classifying them according to \ac{DDoS} traffic class, using FlowMeter technology. Thus, we train \ac{ML} algorithms considering the following traffic classes: Benign, DoS-DNS, DoS-MSSQL, DoS-NetBIOS, DoS-SNMP, DoS-UDP, Syn, TFTP, and UDP-lag which are malign. For training the \ac{ML} models we use the strategy of dividing the dataset 80\% for training and 20\% for testing. In our training strategy, we use the following \ac{ML} algorithms: \ac{KNN}, Random Forest, Decision Tree, and \ac{MLP}. The objective of this experiment is to validate the onboarding SFI2 Architecture block with \ac{ML} capabilities in order to classify whether certain traffic refers to \ac{DDoS} attacks or is benign. For this, we use the training time and accuracy metrics to measure how capable \ac{ML}-Agents are of detecting \ac{DDoS} Attacks against the functional blocks of the architecture.

\subsubsection{Results and Discussion}\label{subsubsection:security_results_and_discussion}

Thus, we train all \ac{ML} algorithms considering the dataset and present their accuracies according to Table~\ref{tab:security_ml_results}. According to Table~\ref{tab:security_ml_results}, we confirmed that \ac{MLP} produces the best accuracy among its peers. However, it requires a significantly higher training time than its peers. On the other hand, \ac{KNN} performed well in terms of accuracy while taking the least amount of time to train.

\begin{table}[htb]
\centering
\caption{Accuracy for \ac{DDoS} Attacks Classification.}
\label{tab:security_ml_results}
\begin{tabular}{lcc}
\hline
\multicolumn{1}{c}{\textbf{Algorithm}} & \textbf{Accuracy} & \textbf{Training Time (seconds)} \\ \hline
\textbf{KNN}                           & 94\%              & 0.026                            \\
\textbf{Random Forest}                 & 92\%              & 4.379                            \\
\textbf{SVM}                           & 93\%              & 36.920                           \\
\textbf{MLP}                           & 95\%              & 542.661                          \\ \hline
\end{tabular}
\end{table}

Furthermore, we dived into behavior \ac{KNN} algorithm to classify \ac{DDoS} Attacks to measure how their accuracy changes depending on the choice of $K$. For this, we performed cross-validation using and made several separations of training and testing to evaluate the model in various sampling scenarios. Thus, we vary the $K$ from $1$ to $30$ and measure its accuracy according to Fig.~\ref{fig:knn_cross_validation}. As this is cross-validation, for each $K$, we obtained average accuracies for ten (10) different executions, and according to Fig.~\ref{fig:knn_cross_validation} with $K = 4$, we obtained the best accuracy of the model, with an average of $0.895$. Subsequently, as the value of $K$ increases, accuracy drops slightly.

Thus, after training the \ac{ML} models, they can be distributed and activated by the \ac{ML}-Agents along the functional blocks of the SFI2 Architecture to empower these blocks of \ac{DDoS} attacks that can compromise or disable their operation. Furthermore, it is imperative to point out that attacks vary over time, requiring improvement of the model that \ac{ML}-Agent uses. After verifying a \ac{DDoS} attack, the functional block can make some decisions regarding the originator of the malicious traffic, such as dropping the traffic from one of the origins.

\subsection{Sustainability}\label{subsec:sustainability}

\begin{figure}[htbp]
    \centering
    \includegraphics[width=.9\columnwidth]{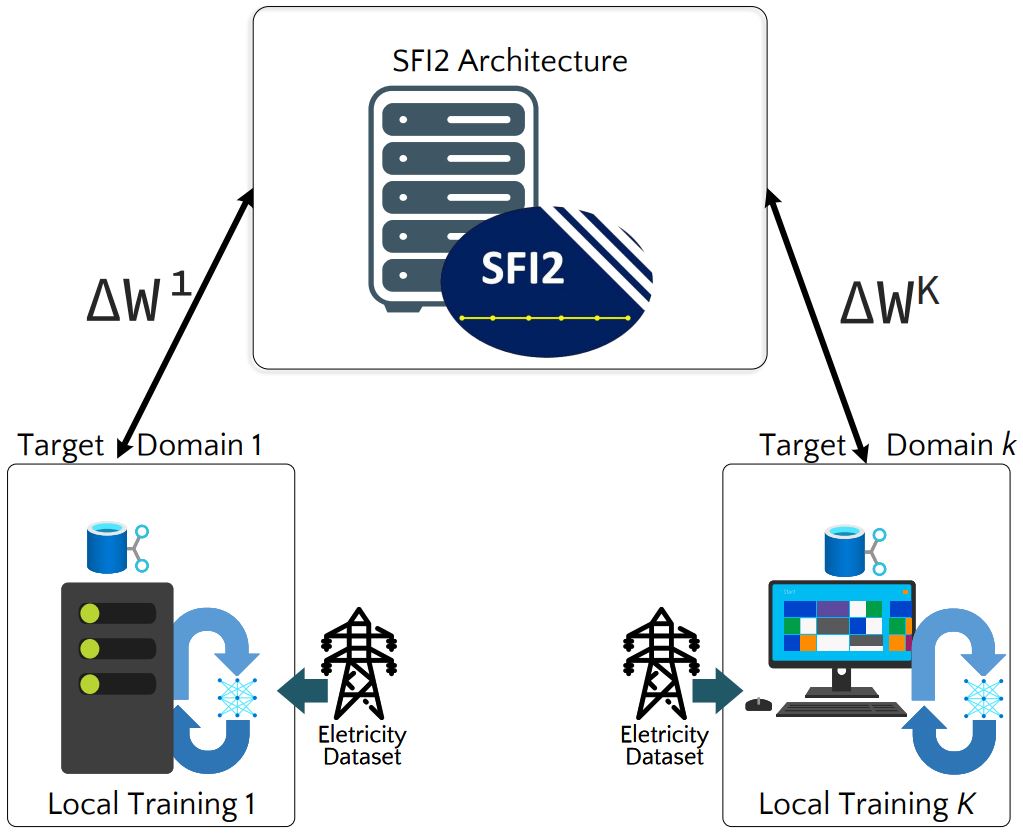}
     \caption{Experimental Setup for Sustainability of SFI2 Reference Architecture.}
    \label{fig:experimental_setup}
\end{figure}

\begin{figure*}[htbp]
\begin{subfigure}{.5\textwidth}
  \centering
  \includegraphics[width=.99\linewidth]{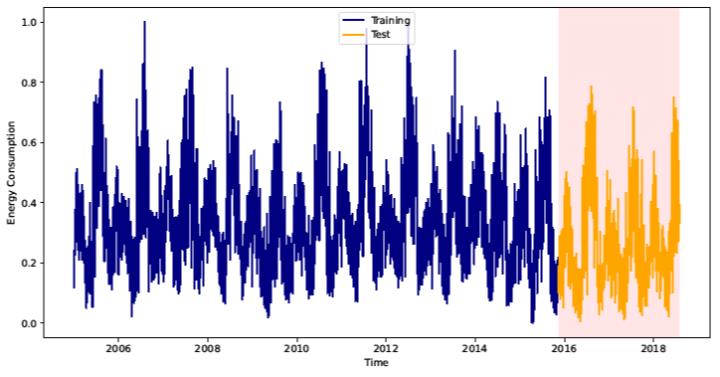}
  \caption{Duquesne Light Company of Pittsburgh.}
  \label{fig:duquesne_pitisburg_dataset}
\end{subfigure}%
\begin{subfigure}{.5\textwidth}
  \centering
  \includegraphics[width=.99\linewidth]{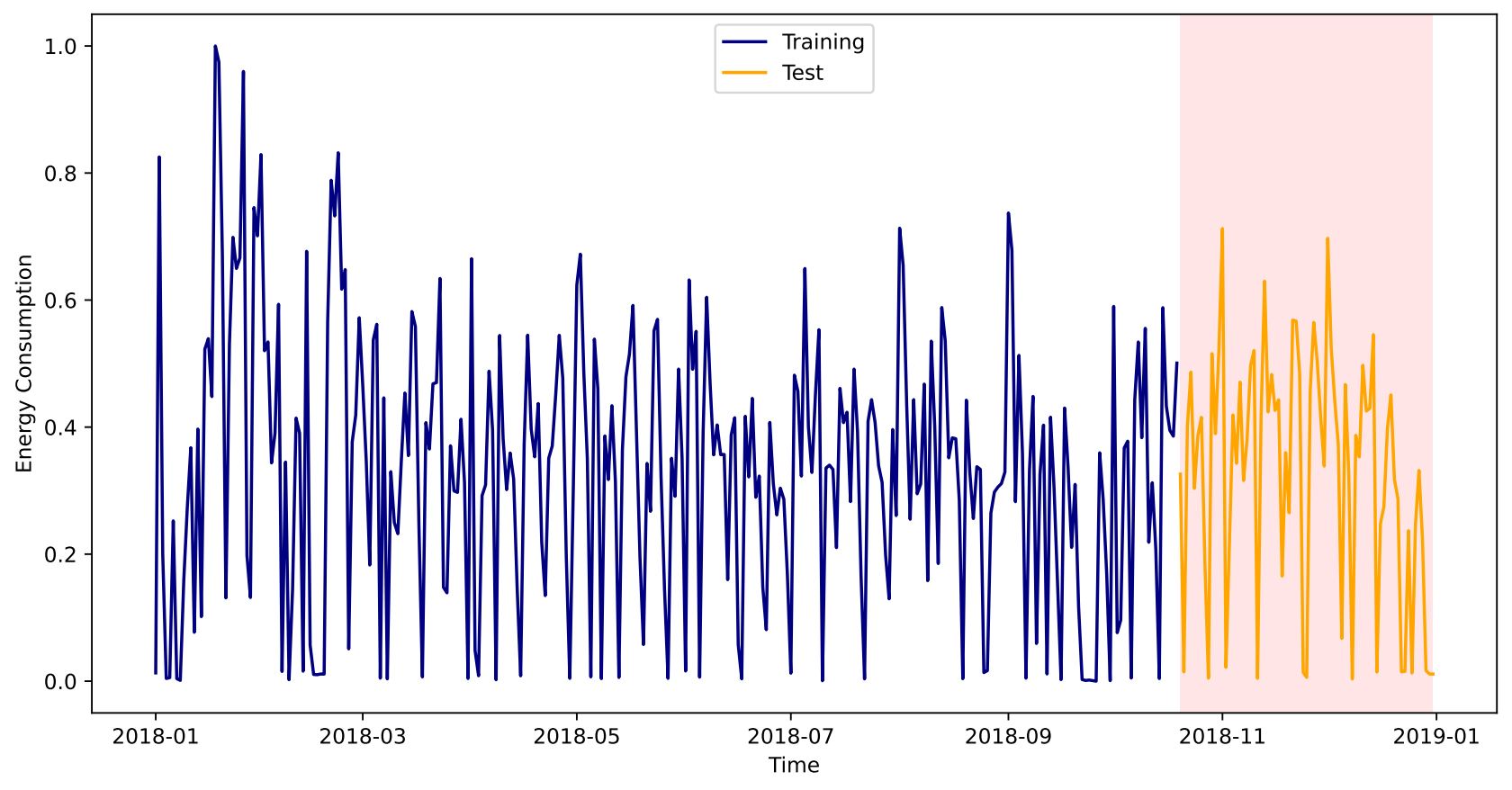}
  \caption{South Korea Steel Industry.}
  \label{fig:south_korea_dataset}
\end{subfigure}
\caption{Electricity Consumption -- Dataset Division.}
\label{fig:dataset_division}
\end{figure*}

We built an experimental scenario to validate the architectural enhancement proposed in SFI2, especially regarding sustainability. Thus, according to Fig.~\ref{fig:experimental_setup}, we consider two different target domains that deliver network slices with specific purposes. These target domains have specific energy sources, so we built a forecasting scenario for electricity consumption in each target domain using the distributed learning paradigm. 

We validate our enhancement suitability with \ac{CNN}, specifically \ac{LSTM} for energy forecasting in each target domain using the SFI2 AI Management block. After training each \ac{ML} model, they are appropriately distributed to the ML-Agent using each domain as shown in Figure~\ref{fig:security_experimental_setup}. After then, the SFI2 architecture can handle energy consumption forecasting.

To validate the forecasting capability of the distributed training, we used the \ac{MSE} metric that measures the difference between the forecast and the actual in a time series. For the proposed experimental scenario, we used the well know neural network the \ac{LSTM}~\cite{sundermeyer2012lstm}.

\subsubsection{Dataset}\label{subsubsec:energy_dataset}

In Fig.~\ref{fig:experimental_setup}, we have a representation of two target domains containing different datasets referring to electricity consumption. The first dataset refers to the Duquesne Light Company's electricity consumption. It contains the multivariate time series of the electricity consumption of the city of Pittsburg, containing the interval December 2005 to January 2018~\cite{duquense}. The second, Steel Industry Energy Consumption Dataset, refers to the electricity consumption of the steel sector in South Korea, containing the record of consumption from January 1, 2018, to December 30, 2018, and the consumption of kWh~\cite{Dua:2019}. 

Thus, the SFI2 resource orchestrator interacts with the two target domains that locally proceed with local data training. We divide the two datasets to 80\% for training and 20\% for testing according to Fig.~\ref{fig:dataset_division}, and we consider a 30-day window for future forecasting in the test set. For reproducibility purposes, we let open the dataset and code artifacts available at \url{https://github.com/romoreira/SFI2-Energy-Sustainability}.

\subsubsection{Results and Discussion}\label{subsubsec:results_and_discussion}

\begin{figure*}[htbp]
\begin{subfigure}{.5\textwidth}
  \centering
  \includegraphics[width=.99\linewidth]{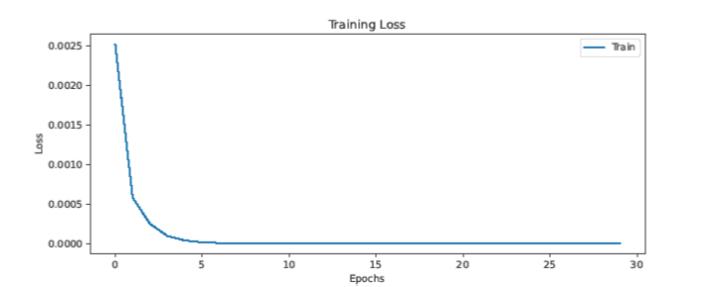}
  \caption{Training Loss Behavior of Target Domain 1.}
  \label{fig:loss_train1}
\end{subfigure}%
\begin{subfigure}{.5\textwidth}
  \centering
  \includegraphics[width=.99\linewidth]{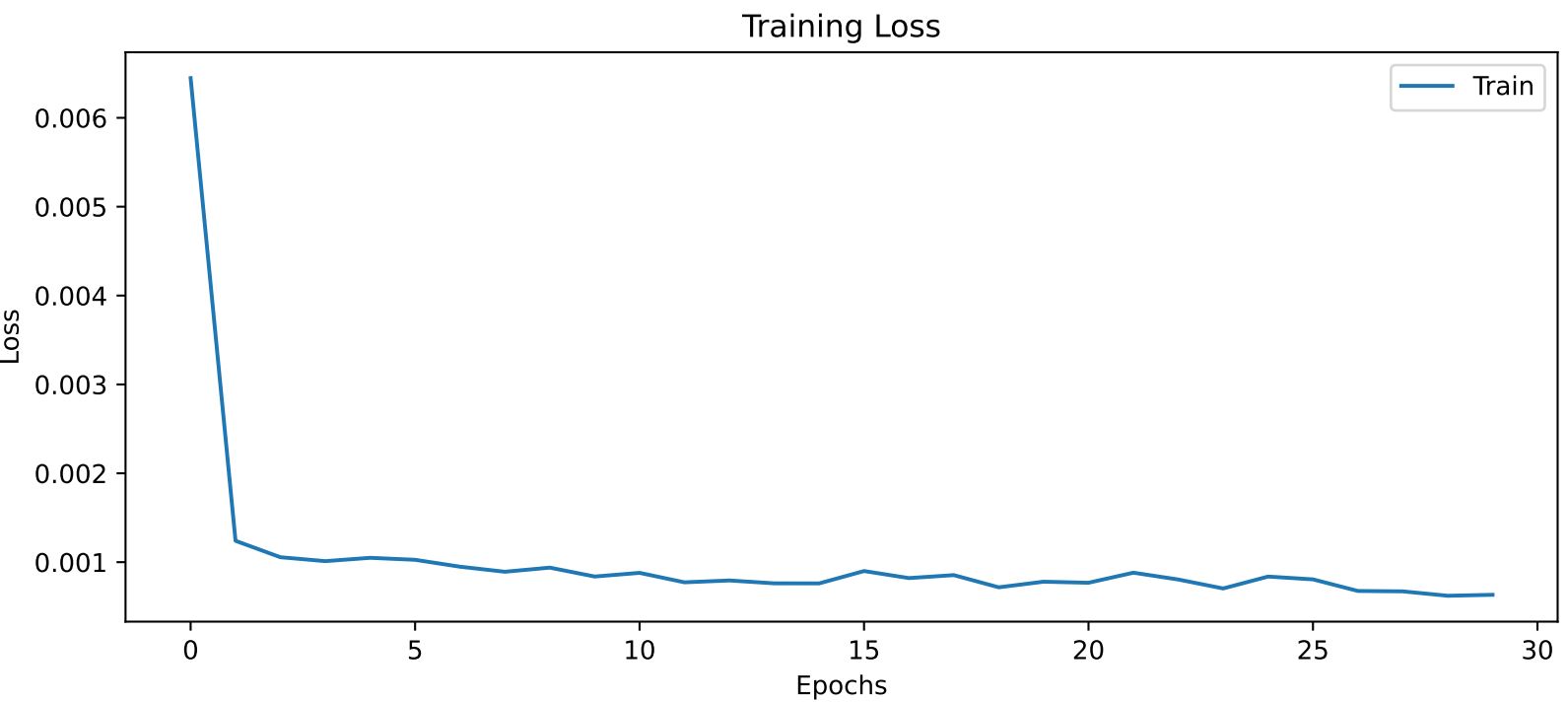}
  \caption{Training Loss Behavior of Target Domain 2.}
  \label{fig:loss_train2}
\end{subfigure}
\caption{Training Loss Behavior of Target Domains}
\label{fig:training_loss}
\end{figure*}

\begin{figure*}
\begin{subfigure}{.5\textwidth}
  \centering
  \includegraphics[width=.99\linewidth]{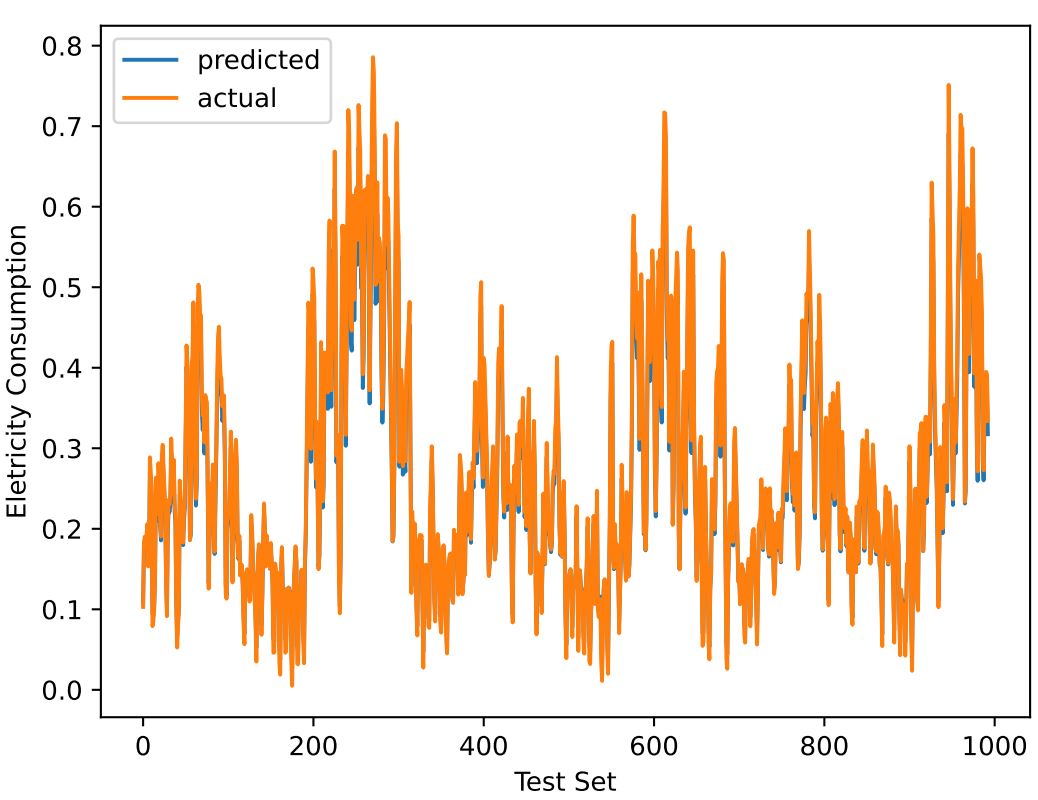}
  \caption{Energy Forecasting Test in Target Domain 1.}
  \label{fig:real_predito1}
\end{subfigure}%
\begin{subfigure}{.5\textwidth}
  \centering
  \includegraphics[width=.99\linewidth]{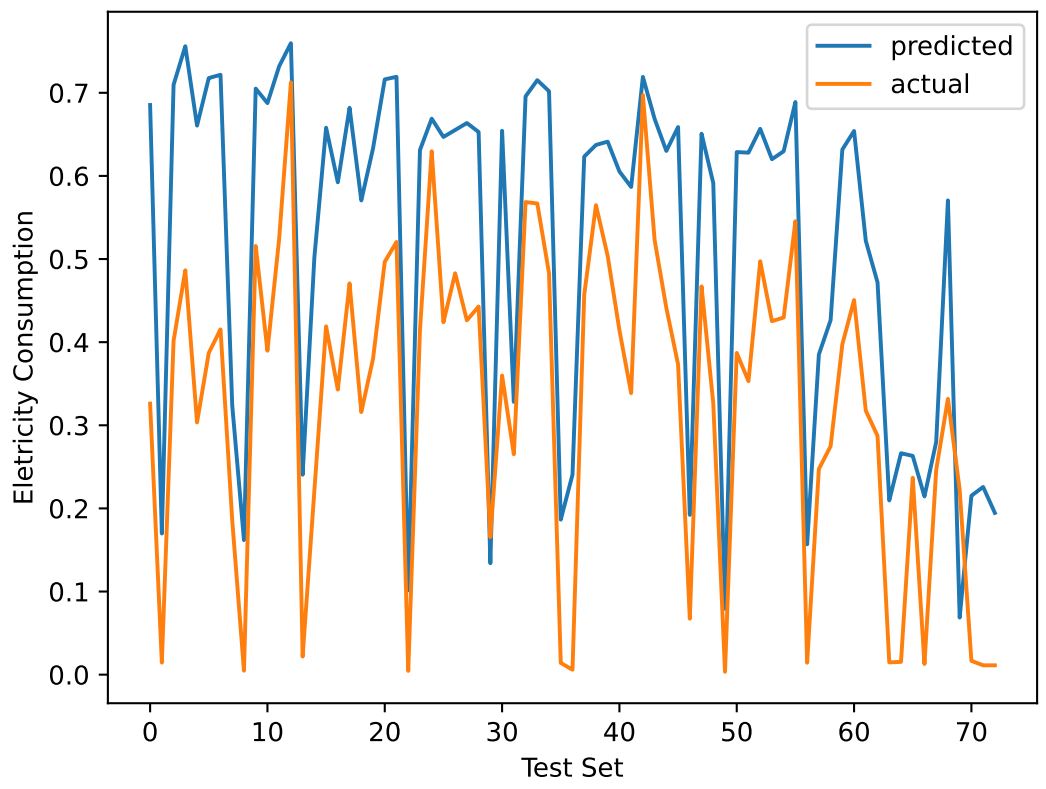}
  \caption{Forecasting Test in Target Domain 2.}
  \label{fig:real_predito2}
\end{subfigure}
\caption{Energy Forecasting Performance Test of SFI2 Reference Architecture.}
\label{fig:forecasting_test}
\end{figure*}

We performed experiments in the proposed scenario to verify the suitability of distributed training for the sustainability use case. Thus, each target domain trained an \ac{CNN} for forecasting time series and reported the model weights to the central model stored by SFI2 Architecture. In addition, we recorded a global \ac{MSE} of 0.0014 on average, which means that the global model's error received the weights of the target domains trained locally with the electricity consumption datasets. Thus, Fig.~\ref{fig:training_loss} depicts the generalization and learning capacity of the model. Given the behavior of the graph, it is suggestive of admitting that the neural network in each target domain was able to learn appropriately due to the descending of the Loss metric.

In addition, we evaluated the ability of the SFI2 Slicing Reference Architecture global model to forecast the electricity consumption of those target domains. We randomly tested the global model using some tested target domains. According to Fig.~\ref{fig:forecasting_test}, distributed learning shows promising results for forecasting electricity consumption in future slicing architectures. We estimated the difference between the actual and the predicted in the forecasting process. The electricity consumption forecast was very accurate, according to its behavior in the time series graph between the actual and predicted curves~Fig.~\ref{fig:forecasting_test}. The error rate is low such that the time series follow similar behavior for both target domains 1 and 2. Our investigation opens up opportunities for further investigation into how sustainability initiatives move toward built-in AI-capabilities in network slicing architectures.

Our sustainability enhancement opens up research opportunities especially considering our distributed artificial intelligence approach. Forecasting target domain energy consumption allows the SFI2 Architecture to ponder its network slicing decisions against the energy constraints of each target domain. Our security-oriented and AI-native architecture fosters a high customization level of slicing architecture.

\section{Final Considerations}
\label{sec:conclusion}

The new SFI2 network slicing reference architecture fills an existing gap resulting from the diversity of NS architecture's target domains by integrating multi-domain and multi-technology experimental network infrastructures. The experimental network integration is enhanced by embedding machine learning native optimizations, energy-efficient slicing, and slicing-tailored security functionalities. An essential architectural design aspect is that although the SFI2 instantiation and deployments were focused on integrating Brazilian experimental testbeds (FIBRE-NG, CloudNEXT, FUTEBOL, FIWARE, 5GINFIRE, and NECOS), the agnostically defined domain and monitoring interfaces (DOM-IM and DOM-MON) allow instantiation to distinct experimental domains. Moreover, the generic embedded enhancements (ML-native optimization, energy efficiency, and tailored security functions) will enable the use of the SFI2 reference architecture in domains other than the experimental ones.

Beyond technical achievements and enhancements, the SFI2 reference architecture significantly benefits the research community by fostering the utilization of currently available and deployed experimental testbeds. The gains are twofold, with enhancements to the current infrastructures and simultaneously allowing the reuse of the infrastructures with additional value. In practice, the achieved gains mean having more extensive resources (VMs, 5G functions, IoT, and others) for experimenters by integrating the infrastructures.

The SFI2 architecture contributes to the network slicing state-of-the-art in various ways. Initially, the SFI2 architecture focuses on integrating experimental networks and allowing the creation, operation, and decommissioning of independent and isolated slices on distinct experimental network domains, fulfilling an existing research gap. These results in reusing available infrastructures, thus reducing capital expenditure (CAPEX) and operational expenditure (OPEX) for the institutions involved with the integrated experimental networks. Additionally, slicing in the SFI2 architecture incorporates significant global values and principles by adopting a sustainability-aware dynamic resource discovery and leasing for multi-domain experimental networks.

Besides, the modular and agnostic design characteristics of the SFI2 main interface modules with the experimental domains through DOM-IM and DOM-MON, together with the embedded ML-native optimizations and slice-tailored security functions, push the network slicing paradigm to the level of functionalities required by current and future users and applications.

Several research opportunities lie ahead using the proposed architecture. First, it is necessary to extend SFI2 to support a more significant number of experimental and possibly operational networks. We expect to face new challenges as we include new domains and technologies in a multi-domain scenario. Second, the ML selection for each optimization task is challenging, extending and evaluating different approaches. In this context, the training datasets, i.e., data collection and labeling for each domain and technology, should be investigated and enhanced for slice-as-a-service.

Finally, the SFI2 architecture allows for advancing new research challenges. For example, providing comprehensive security tests to prevent zero-day attacks is critical. Moreover, tests of reconfiguration of slices considering the elasticity of resource allocation according to demand and sustainable approaches are desirable for future architecture versions. In resource allocation, testing and guaranteeing the isolation of network slices created as services is also essential.

\bibliographystyle{IEEEtran}
\typeout{}
\bibliography{ref.bib}

\vskip -2\baselineskip plus -2fil

\begin{IEEEbiography}
    [{\includegraphics[width=1in,height=1.25in,clip,keepaspectratio]{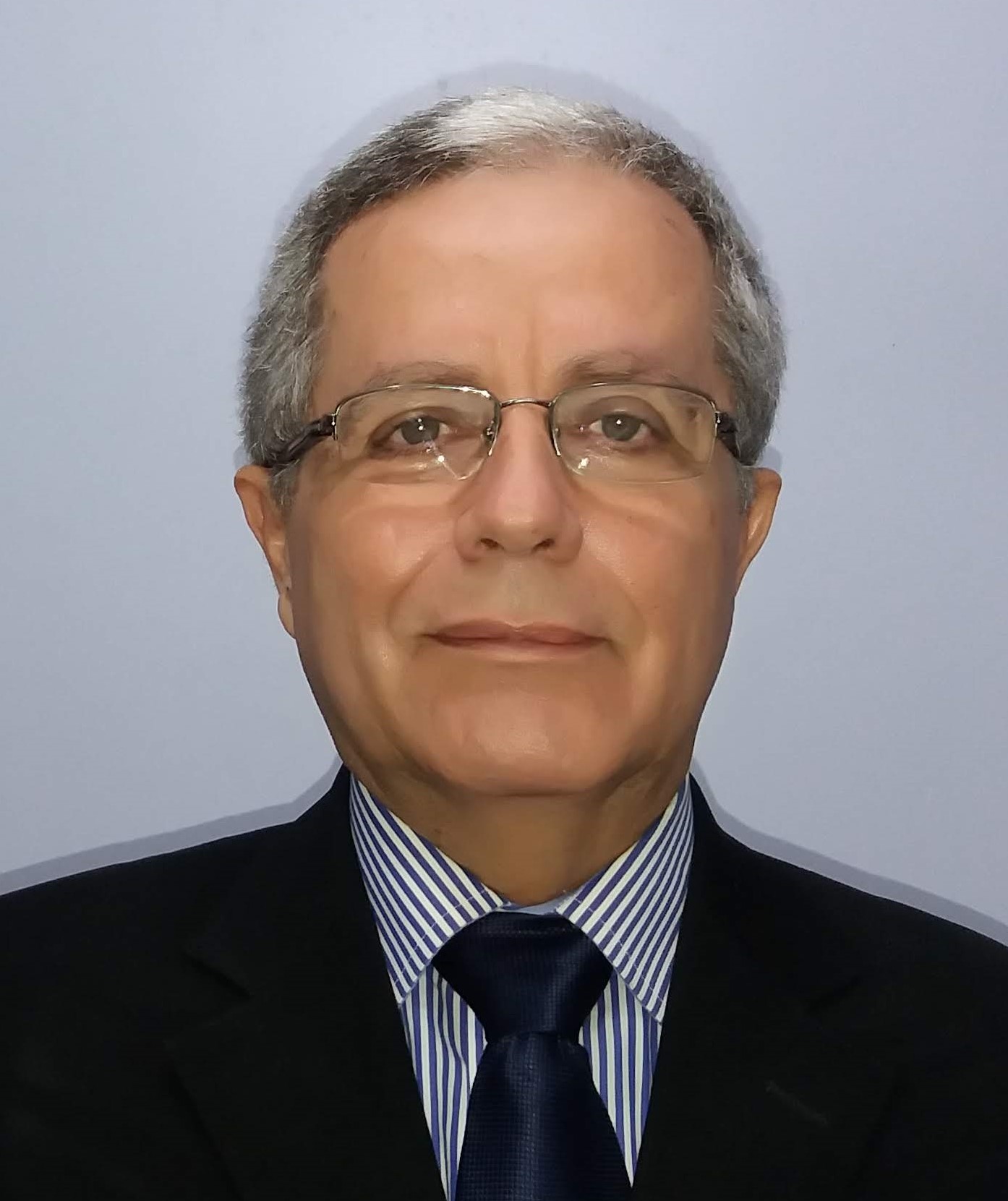}}]{\bf{JOBERTO S. B. MARTINS} (Life Senior Member, IEEE) -} Ph.D. in Computer Science at Université Pierre et Marie Curie - UPMC, Paris (1986), PosDoc at ICSI - Berkeley University (1995), and PosDoc Senior Researcher at Paris Saclay University - France (2016). International Professor at Hochschule für Technik und Wirtschaft des Saarlandes - HTW (Germany) (since 2004) and Université d'Evry (France). Full Professor at Salvador University (UNIFACS) on Computer Science, Director of NUPERC and IPQoS research groups with research interests in Network Slicing, Resource Allocation, Machine Learning, Software-defined Networking, Smart Grid, and Smart City. He is a key speaker, teacher, and invited lecturer at various international congresses and companies in Brazil, USA, and Europe.
\end{IEEEbiography}

\vskip -2\baselineskip plus -1fil

\begin{IEEEbiography}
    [{\includegraphics[width=1in,height=1.25in,clip,keepaspectratio]{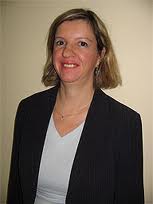}}]{\bf{TEREZA C. CARVALHO} -} Associate Professor of Escola Politécnica – University of São Paulo (USP) and visiting professor at Université Paris 1 Panthéon-Sorbonne. She is the founder and general coordinator of LASSU (Laboratory of Sustainability on ITC) since 2010 and CEDIR-USP (Center for Reuse and Discard of Informatics Residuals) since 2009. She is a former assessor of CTI – USP (Information Technology Coordination) during 2010-2013 and the director of CCE-USP (Electronic Computing Center) during 2006-2010. She is a Sloan Fellow 2002 from MIT (Massachusetts Institute of Technology). She has coordinated international and national R\&D projects since 2000 in Green Computing, Cloud Computing, IT Energy Efficiency, IT Governance, Digital Technologies applied to the Amazon Production Chains, WEEE (Waste Electrical and Electronic Equipment), Future Internet, Scientific DMZ, and Security. She holds several international patents.
\end{IEEEbiography}

\vskip -2\baselineskip plus -1fil

\begin{IEEEbiography}
    [{\includegraphics[width=1in,height=1.25in,clip,keepaspectratio]{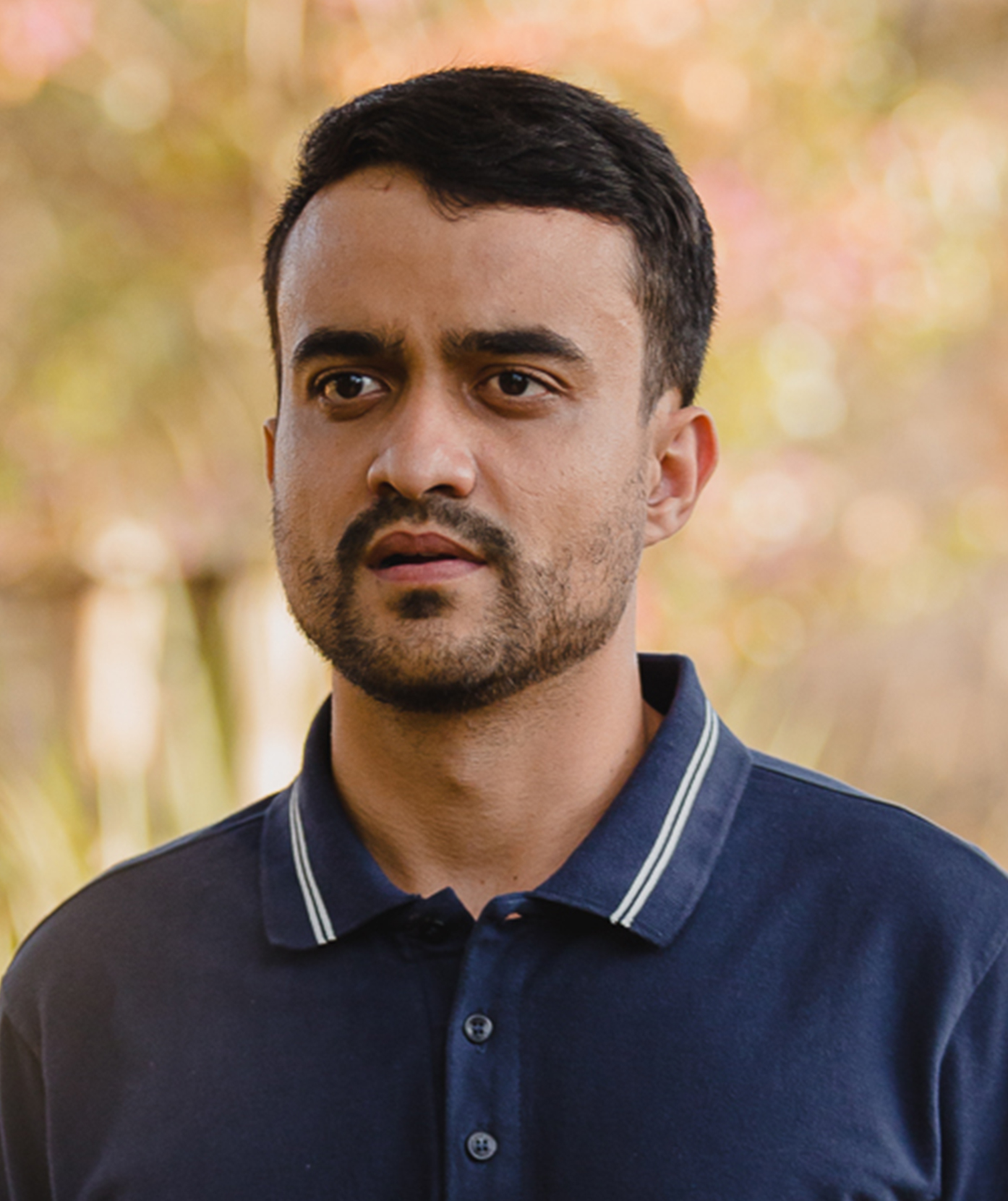}}]{\bf{RODRIGO MOREIRA} -} is a Professor at the Federal University of Viçosa and received his B.S. degree from the Federal University of Viçosa and his M.S. degree from the Federal University of Uberlândia, Brazil, in 2014 and 2017 respectively. Also, he received in 2021 a Ph.D. degree in Computer Science at the Federal University of Uberlândia. His research interests include future internet, quality of service, cloud computing, network function virtualization, software-defined networking, and edge computing.
\end{IEEEbiography}

\vskip -2\baselineskip plus -1fil

\begin{IEEEbiography}
    [{\includegraphics[width=1in,height=1.25in,clip,keepaspectratio]{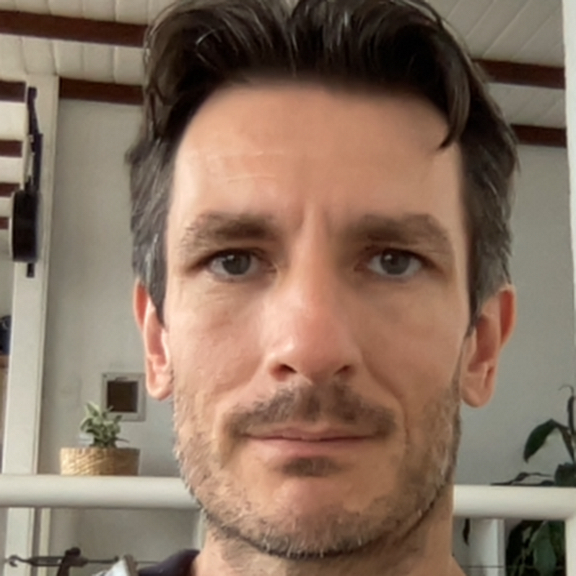}}]{\bf{CRISTIANO BONATO BOTH} -} is a professor of the Applied Computing Graduate Program at the University of Vale do Rio dos Sinos (UNISINOS), Brazil. He coordinates research projects funded by H2-1.52-1.5 EU-Brazil, CNPq, FAPESP, and RNP. His research focuses are on wireless networks, next-generation networks, softwarization, and virtualization technologies for telecommunication networks.

\end{IEEEbiography}

\vskip -2\baselineskip plus -1fil

\begin{IEEEbiography}
    [{\includegraphics[width=1in,height=1.25in,clip,keepaspectratio]{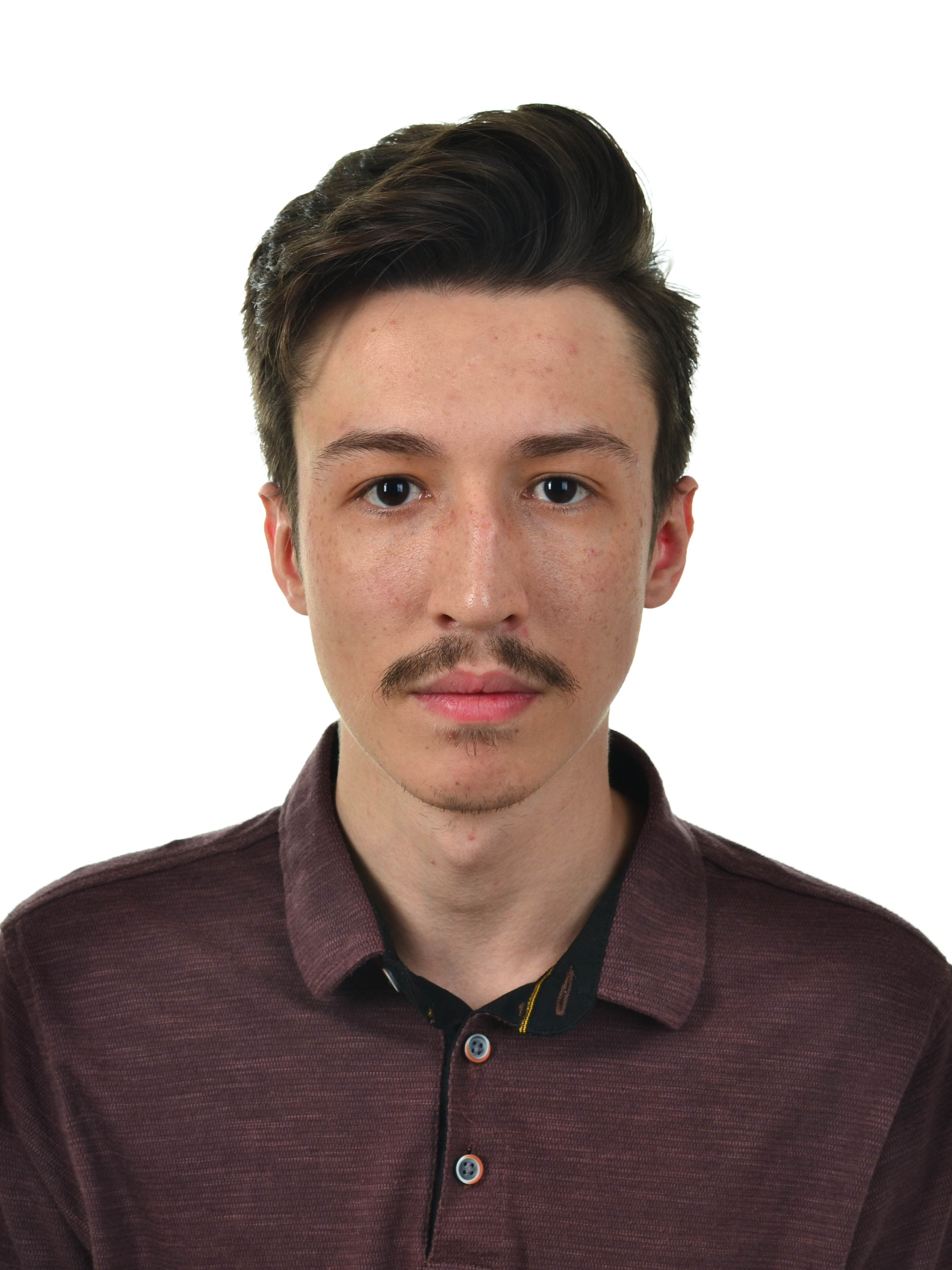}}]{\bf{ADNEI DONATTI} -}
 Ph.D. student at the University of São Paulo (USP) working with resource allocation for Network Slicing on the SFI2 project. Received the M.Sc. degree in Applied Computing and Bachelor of Technology from Santa Catarina State University (UDESC). Interested in cloud computing, distributed systems, and computer networks.
\end{IEEEbiography}

\vskip -2\baselineskip plus -1fil

\begin{IEEEbiography}
    [{\includegraphics[width=1in,height=1.25in,clip,keepaspectratio]{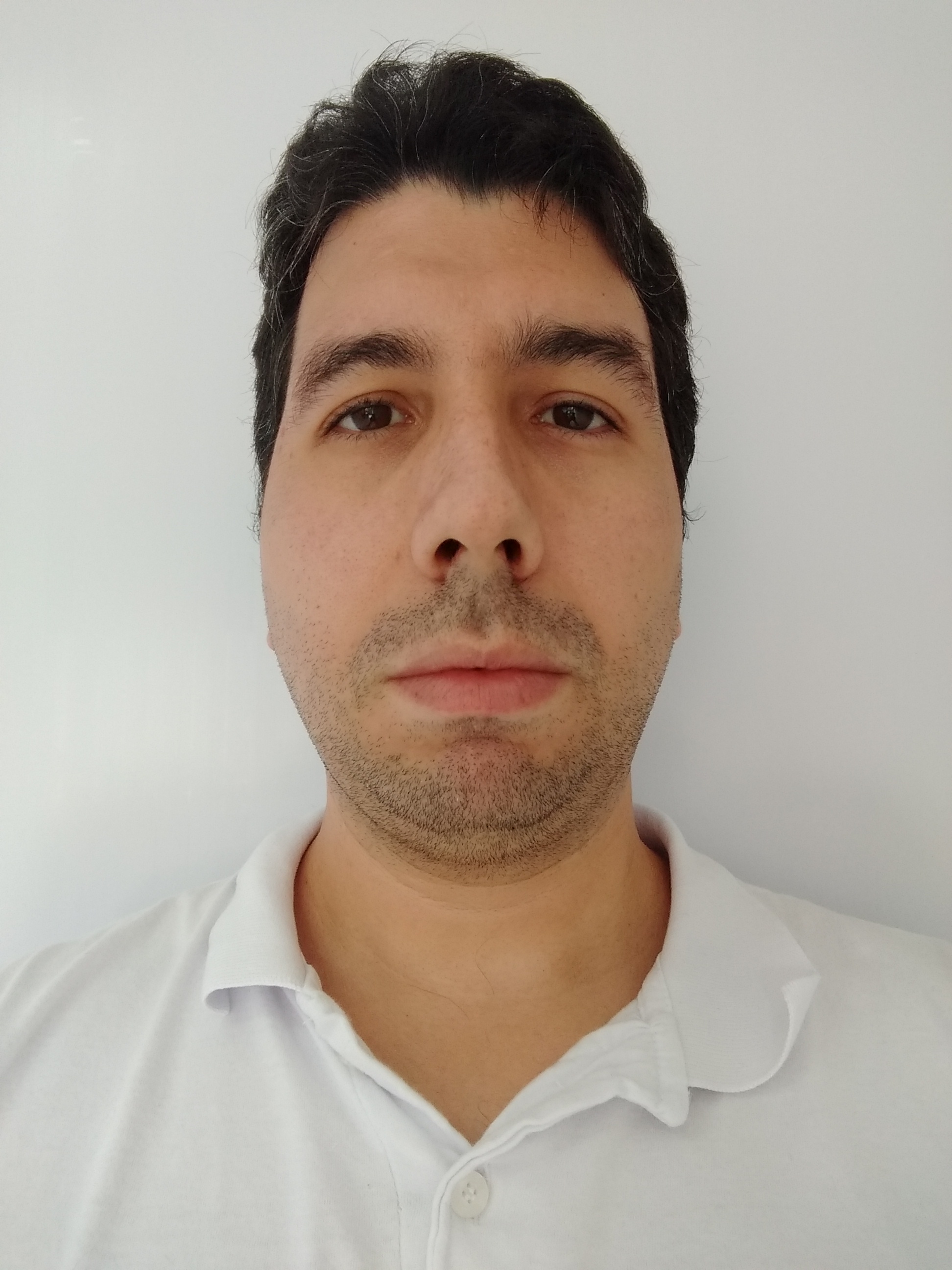}}]{\bf{João H. Corrêa} -} is currently an assistant professor at the Federal University of Ceará (UFC). He received his Ph.D. in Computer Science in 2021 at the Federal University of Espírito Santo (UFES), his M.Sc. degree in Computer Science from the Federal University of Paraíba (UFPB) in 2017, and his B.Sc degree in Computer Networks from the Federal Institute of Paraíba (IFPB) in 2015. His current research interests include Computer Networks, Software-Defined Networks, Network Security, and Cloud Computing.
\end{IEEEbiography}

\vskip -2\baselineskip plus -1fil

\begin{IEEEbiography}
    [{\includegraphics[height=1.25in,clip,keepaspectratio]{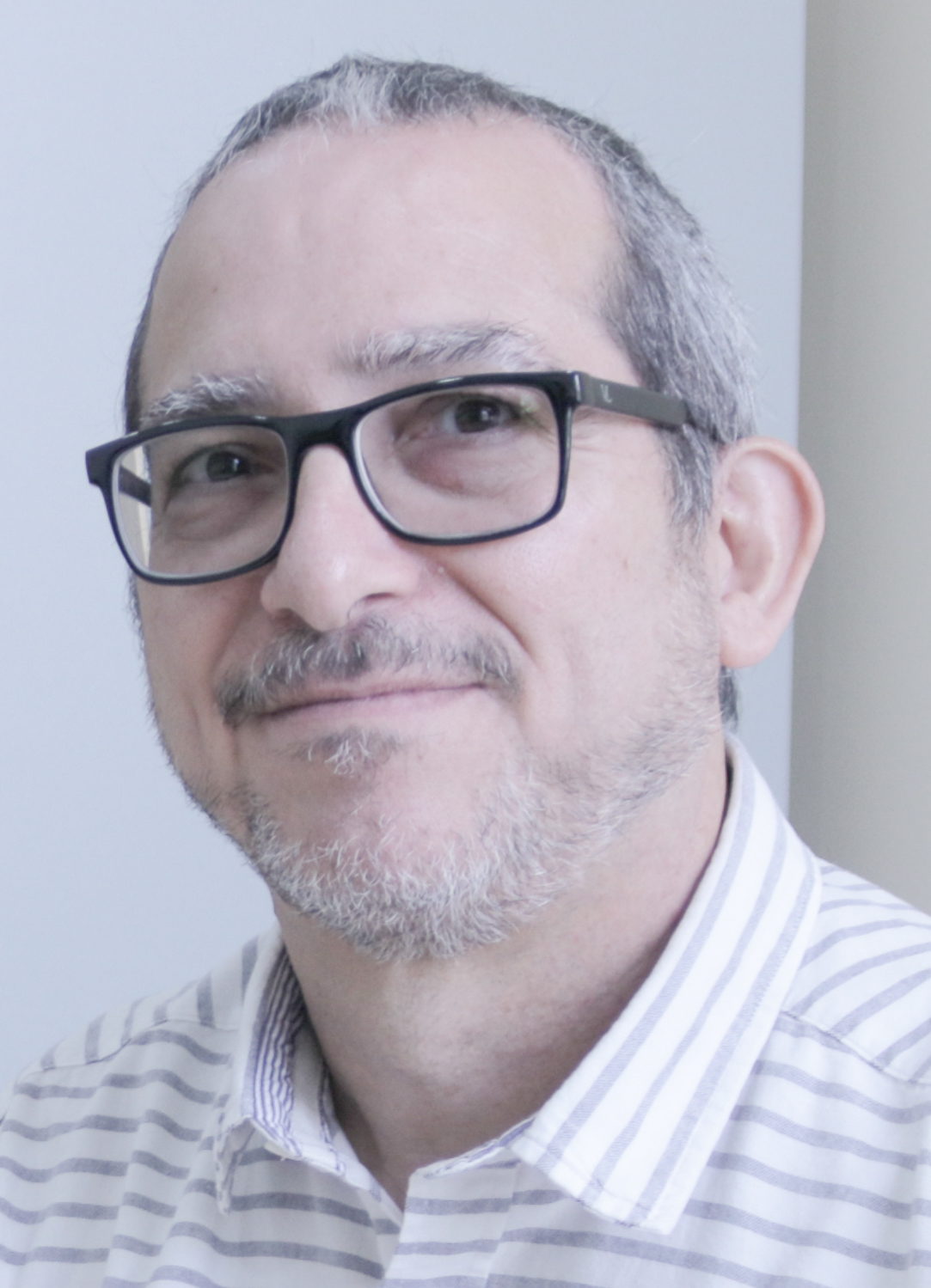}}]{\bf{JOSÉ A. SURUAGY}  (Member, IEEE) -} graduated in Electric Engineering at the Federal University of Pernambuco -- UFPE (1979), a master's in Electric Engineering from the University of São Paulo -- USP (1982), and a Ph.D. in Computer Science from the University of California at Los Angeles -- UCLA (1990). Has experience in Computer Science, focusing on Telecomputing, acting on the following subjects: computer networks, measurement, quality of service, and performance evaluation. In 2014, received the “Computer Network Brazilian Symposium Distinguished Premium” for his scientific contributions in the areas of computer networks and distributed systems, for his participation in the Symposium activities, and for his services that benefitted that community. Currently, he is a full professor and researcher at Centro de Informática (UFPE).
\end{IEEEbiography}

\vskip -2\baselineskip plus -1fil

\begin{IEEEbiography}
    [{\includegraphics[width=1in,height=1.25in,clip,keepaspectratio]{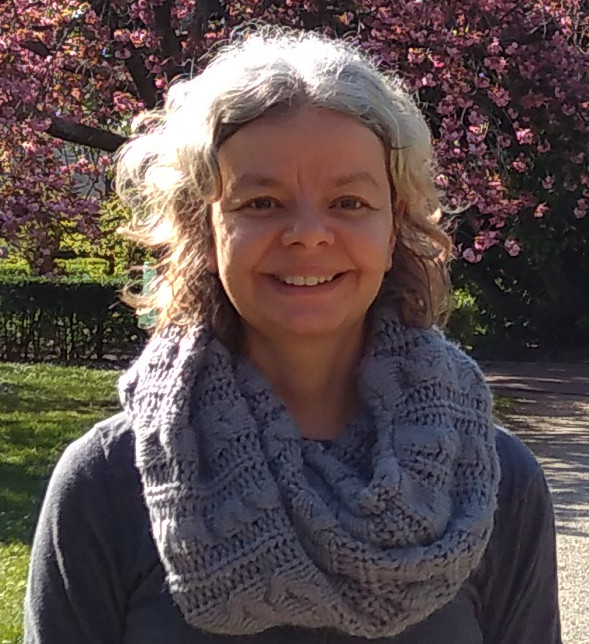}}]{\bf{Sand L. Corrêa} -}
 Sand L. Corrêa received a bachelor's degree in Computer Science from the Federal University of Goiás (UFG), in 1994. In 1997, she received an M.Sc. degree in Computer Science from the State University of Campinas (Unicamp). She received a D.Sc. degree in Informatics from the Pontifical Catholic University of Rio de Janeiro (PUC-Rio), in 2011. Since 2010, she is an associate professor at the Institute of Informatics at UFG.
\end{IEEEbiography}

\vskip -2\baselineskip plus -1fil

\begin{IEEEbiography}
    [{\includegraphics[width=1in,height=1.25in,clip,keepaspectratio]{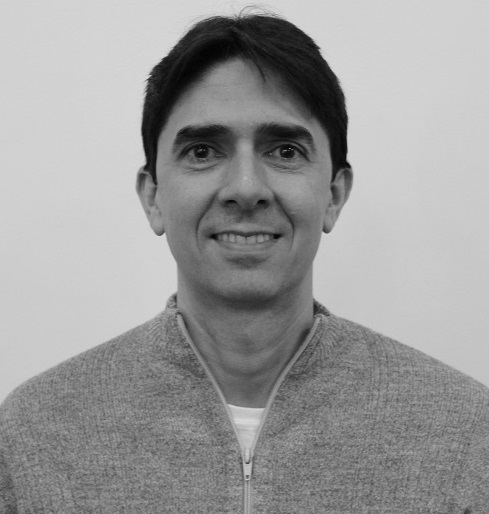}}]{\bf{Antonio J. G. Abelem} (Member, IEEE) -}
 Antônio  Abelém is full Professor in the Computer Science Faculty at the Federal University of Pará (UFPA) in Belém, capital of the Brazilian state of Pará. He holds a Ph.D. in Computer Science from the Catholic University of Rio de Janeiro (PUC-Rio, 2003). He is an associate and member of the Board of the Brazilian Computer Society (SBC). He coordinates the Research Laboratory in Computer Network and Multimedia Communication (GERCOM) from which participates and coordinates several national and international projects. In 2020, he was visiting professor at the University of Massachusetts (UMass), in Amherst-MA, where he carried out research on quantum networks. His research is focused on the field of networking.
\end{IEEEbiography}

\vskip -2\baselineskip plus -1fil

\begin{IEEEbiography}
    [{\includegraphics[width=1in,height=1.25in,clip,keepaspectratio]{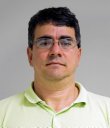}}]{\bf{Moisés R. N. Ribeiro} -}
received  a  B.Sc.  degree  in  electrical  engineering from the Instituto Nacional de Telecomunicações, Brazil, in 1992, the M.Sc.degree in telecommunications from the Universidade Estadual de Campinas, Brazil, in 1996, and a Ph.D. degree from the University of Essex, U.K., in 2002.  In  1995,  he  joined  the  Department  of  Electrical  Engineering,  the Federal University of Espírito Santo. He was a Visiting Professor at the Photonics and Networking Research Laboratory, at Stanford University from 2010 to 2011. His research interests include fiber optic communication and sensor devices, systems, and networks.
\end{IEEEbiography}

\vskip -2\baselineskip plus -1fil

\begin{IEEEbiography}
    [{\includegraphics[width=1in,height=1.25in,clip,keepaspectratio]{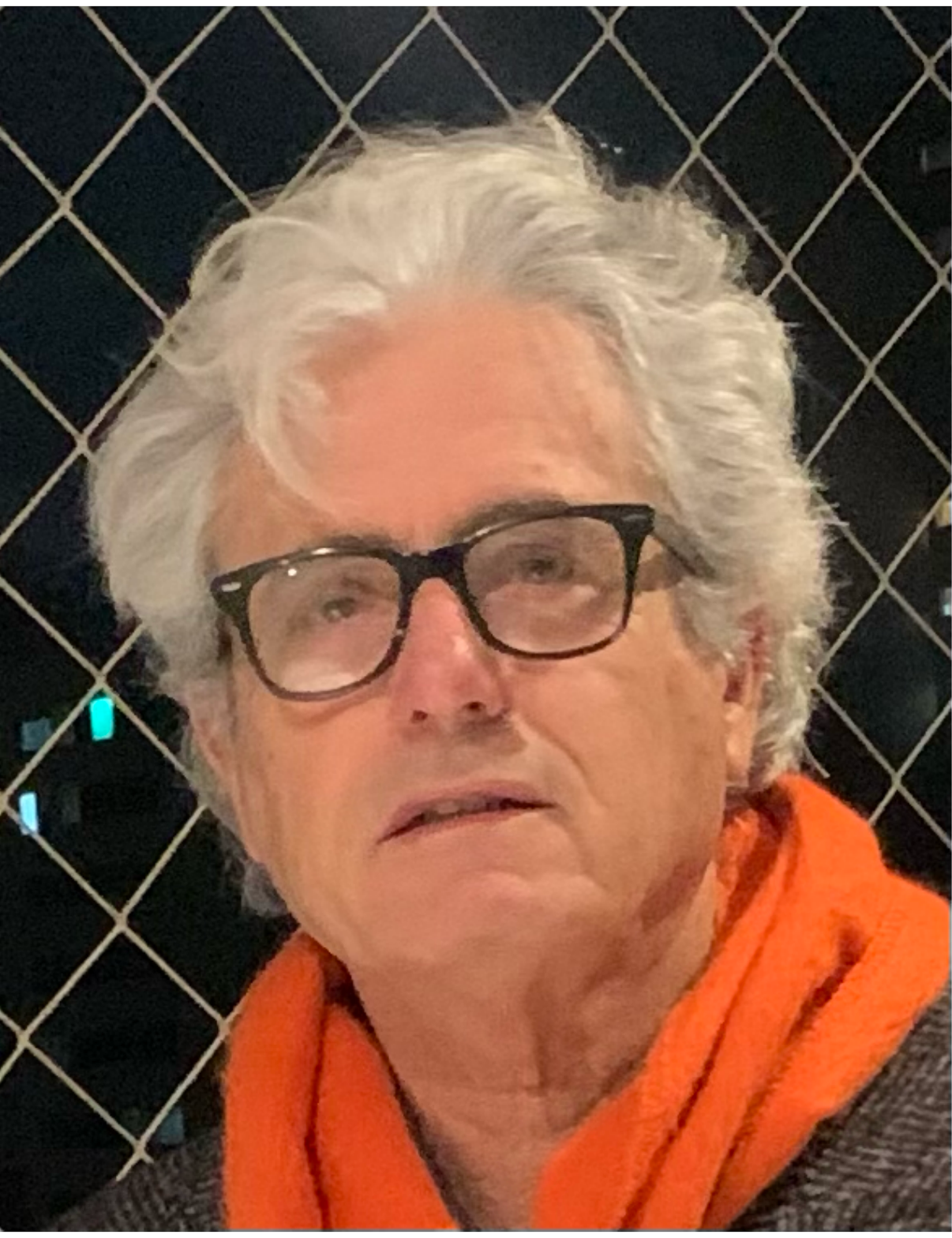}}]{\bf{JOSÉ MARCOS S. NOGUEIRA} (Member, IEEE) -}

 is a full professor (nowadays collaborator retired) of Computer Science at Universidade Federal de Minas Gerais (UFMG), Brazil. His areas of interest and research include (computer, management, wireless sensor, mobile and vehicular, programmable, and security) networks. He received a BS degree in Electrical Engineering and an MS degree in Computer Science from UFMG (1979), and a Ph.D. degree in Electrical Engineering/Computing from the University of Campinas (1985). He held a Post-Doctoral position at the UBC, (Canada), and had a Sabbatical year at the universities Pierre et Marie Curie (Paris 6) and Evry Val D‘Essonne, (France). He chaired/headed many activities at UFMG (department, graduate, undergraduate) where he has been for few months dean and vice-dean. Heads the Brazilian multi partner's project PORVIR-5G. Served as chair/co-chair in several OCs and TPCs. Is a member of diverse editorial boards.   In 2015, he was awarded the computer network community highlight award.
\end{IEEEbiography}

\vskip -2\baselineskip plus -1fil

\begin{IEEEbiography}
    [{\includegraphics[width=1in,height=1.25in,clip,keepaspectratio]{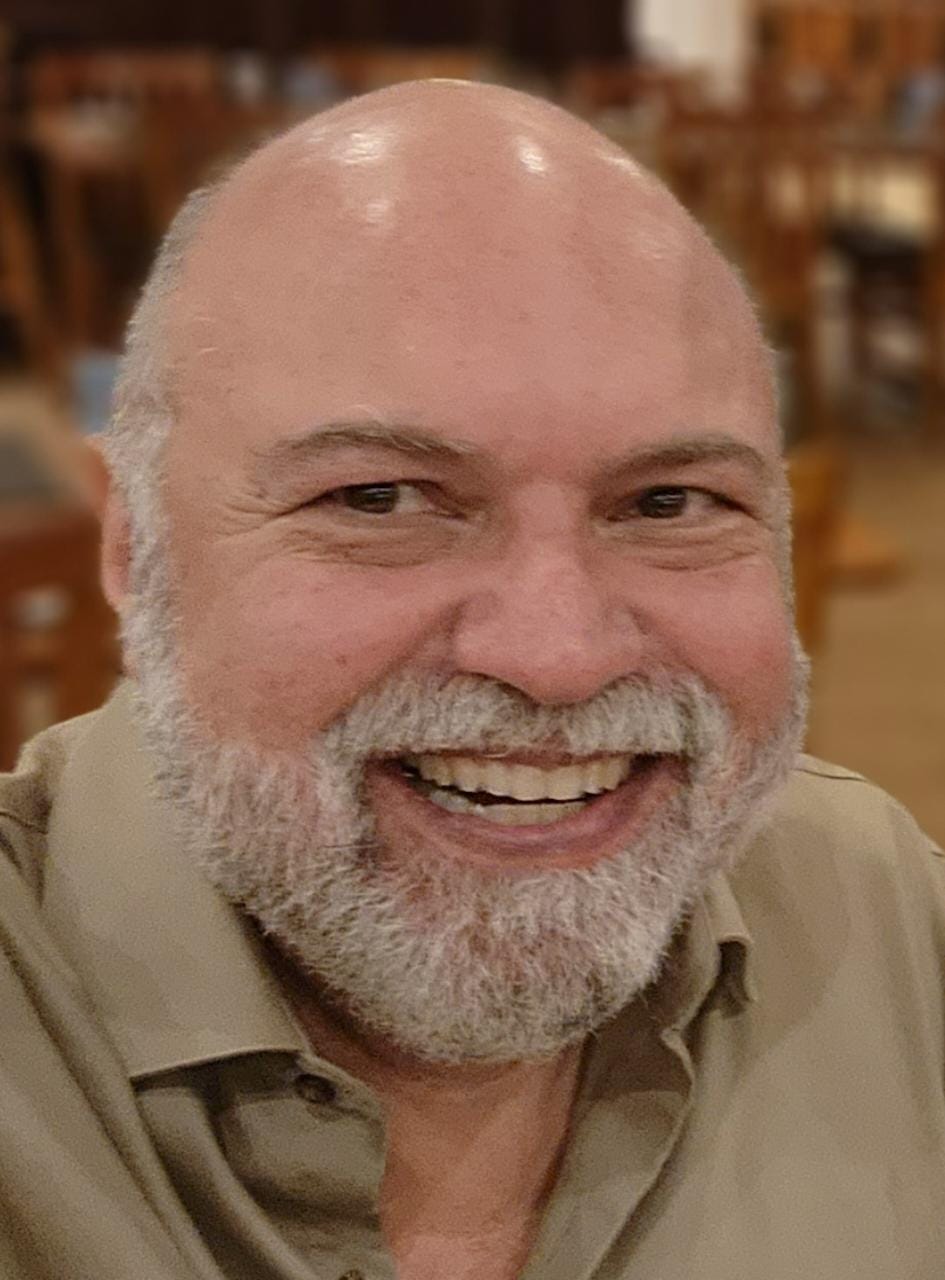}}]{\bf{Luiz C. S. Magalhães} -}is a Professor at the Telecommunications Engineering Department at Unversidade Federal Fluminense (UFF), Niterói, RJ.  He has a PhD in Computer Science from the University of Illinois at Urbana-Champaign (2005) and an MSc in Informatics from the Pontifícia Universidade Católica do Rio de Janeiro (PUC-RJ) (1993)  His main areas of interest are mobile computing, transport protocols, and large scale infrastructure.  
\end{IEEEbiography}

\vskip -2\baselineskip plus -1fil

\begin{IEEEbiography}
    [{\includegraphics[width=1in,height=1.25in,clip,keepaspectratio]{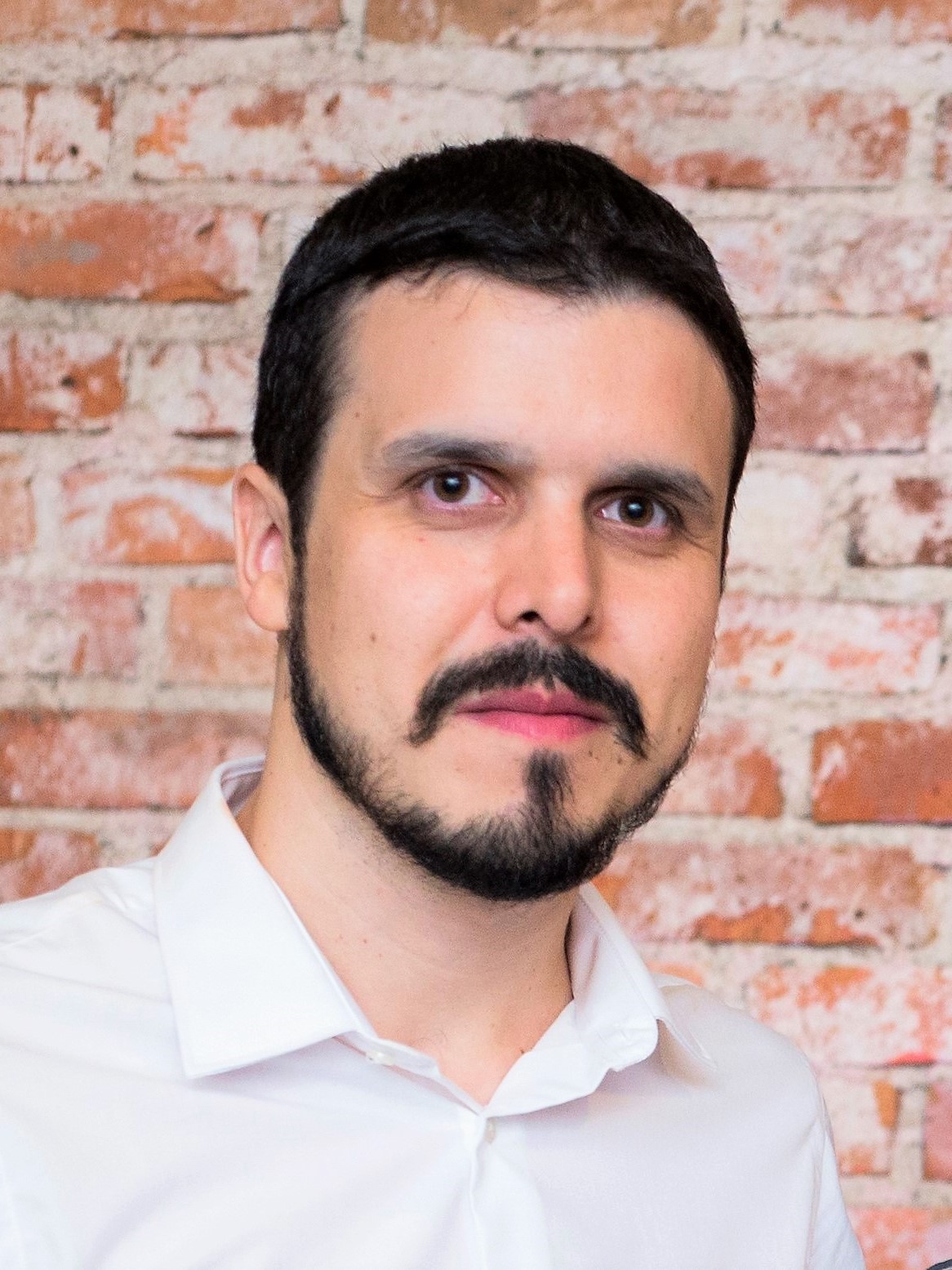}}]{\bf{Juliano Wickboldt} -} is an associate professor at the Federal University of Rio Grande do Sul (UFRGS) in Brazil. He holds both M.Sc. (2010) and Ph.D. (2015) degrees in computer science from UFRGS. Juliano was an intern at NEC Labs Europe in Heidelberg, Germany for one year between 2011 and 2012. In 2015, Juliano was a visiting researcher at the Waterford Institute of Technology in Ireland. His research interests include softwarized networking, IoT, and 5G technologies.
\end{IEEEbiography}

\vskip -2\baselineskip plus -1fil

\begin{IEEEbiography}
    [{\includegraphics[width=1in,height=1.25in,clip,keepaspectratio]{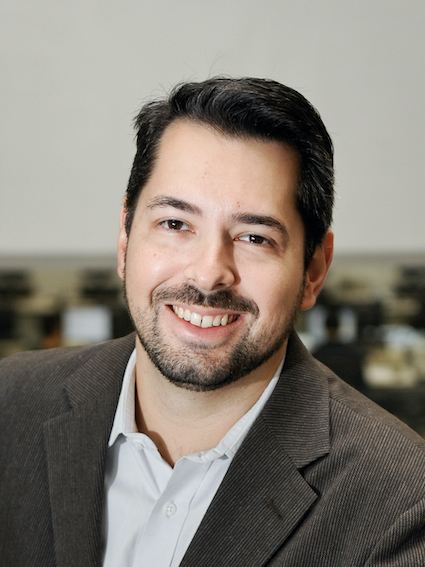}}]{\bf{Tiago C. Ferreto} -} is an associate professor of Computer Science at the Pontifical Catholic University of Rio Grande do Sul (PUCRS), Brazil. He received his Ph.D. in Computer Science from the Computer Science Department, PUCRS, Brazil (2010) with a Sandwich Ph.D. Internship at TU-Berlin, Germany (2007-2008). He is currently the head of GRIN (Research Group on Networking, Infrastructure, and Cloud Computing) and LAD (High-Performance Computing Laboratory) at PUCRS. His research interests include Cloud and Edge Computing, IT Infrastructure Management, Computer Networks, High-Performance Computing, and Big Data.
\end{IEEEbiography}

\vskip -2\baselineskip plus -1fil

\begin{IEEEbiography}
    [{\includegraphics[width=1in,height=1.25in,clip,keepaspectratio]{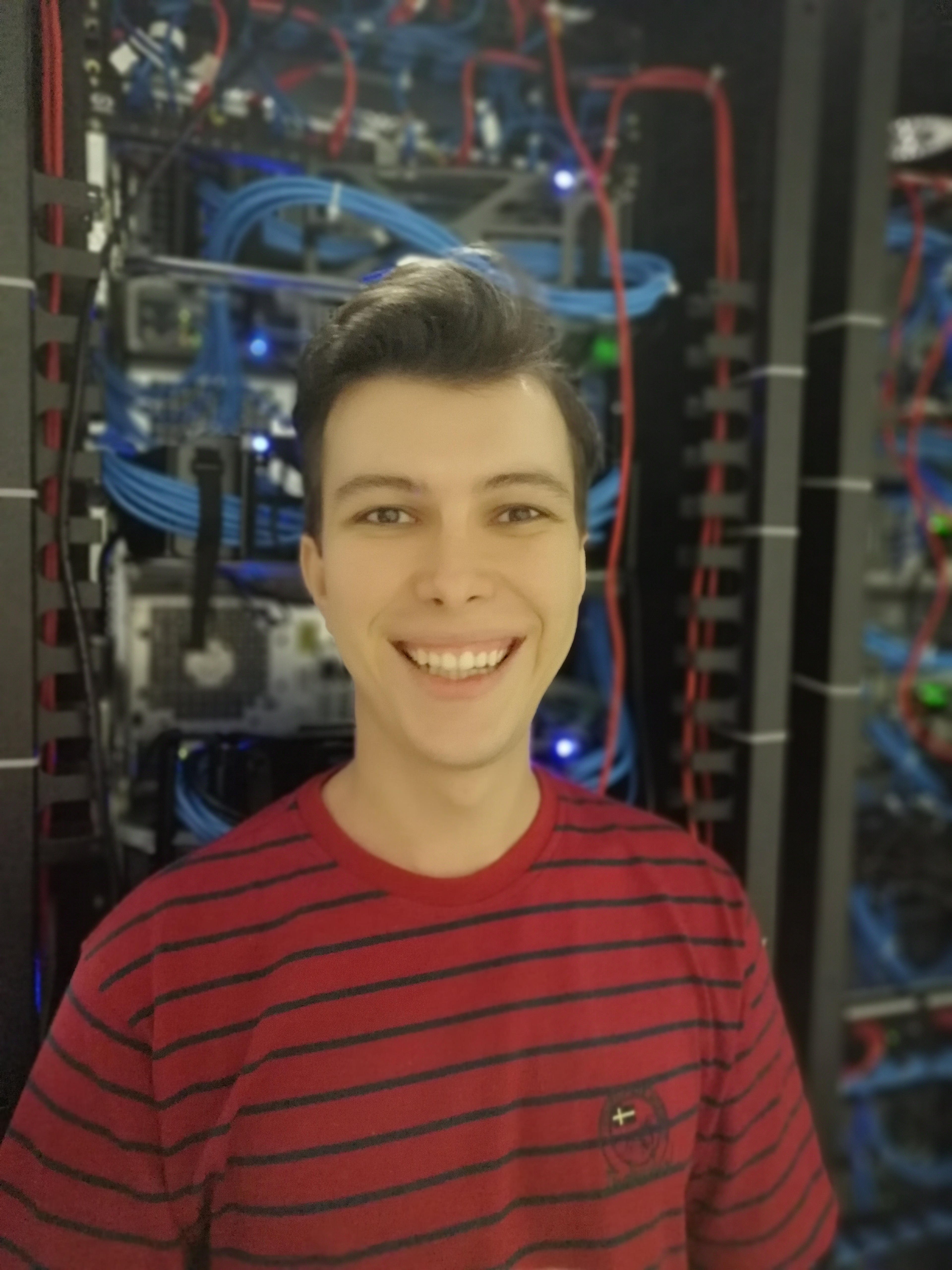}}]{\bf{Ricardo Mello} -} is an assistant professor in the Electrical Engineering Department at the Federal University of Espírito Santo (UFES, Brazil). He received his B.Sc. (2015), his M.Sc. (2018), and his Ph.D (2020) degrees in Electrical Engineering from UFES. His Ph.D. thesis dealt with practical aspects of cloud robotics research and his research resulted recently in a book published in Springer's STAR series. His research interests include cloud robotics, networked systems, autonomous systems, and human–robot interaction.
\end{IEEEbiography}

\vskip -2\baselineskip plus -1fil

\begin{IEEEbiography}
    [{\includegraphics[width=1in,height=1.25in,clip,keepaspectratio]{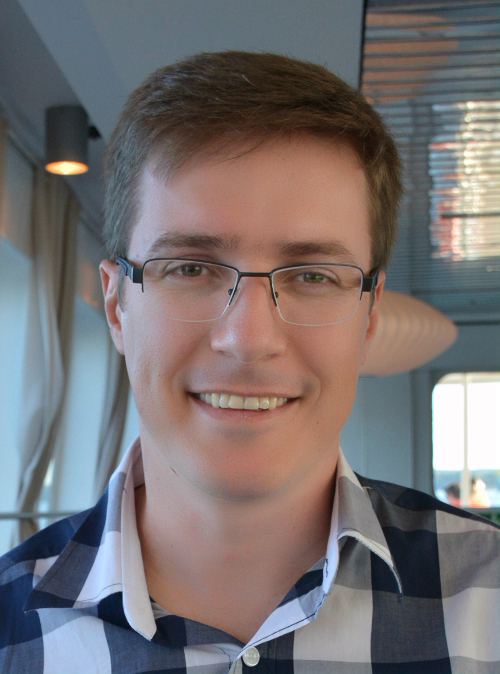}}]{\bf{Rafael Pasquini} (Senior Member, IEEE) - } received his M.Sc. and Ph.D. degrees in computer engineering from the State University of Campinas in 2006 and 2011, respectively. From 2015 to 2017 he was a Visiting Researcher in the Department of Network and Systems Engineering (NSE) at KTH Royal Institute of Technology. Since 2011 he has been an Associate Professor and leads the Distributed Systems and Networks (DSN) research group at the Department of Computer Science of the Federal University of Uberlândia. His research interests include network management, slicing of softwarized infrastructures, machine learning, cloud computing, and software-defined networks. 
\end{IEEEbiography}

\vskip -2\baselineskip plus -1fil

\begin{IEEEbiography}
    [{\includegraphics[width=1in,height=1.25in,clip,keepaspectratio]{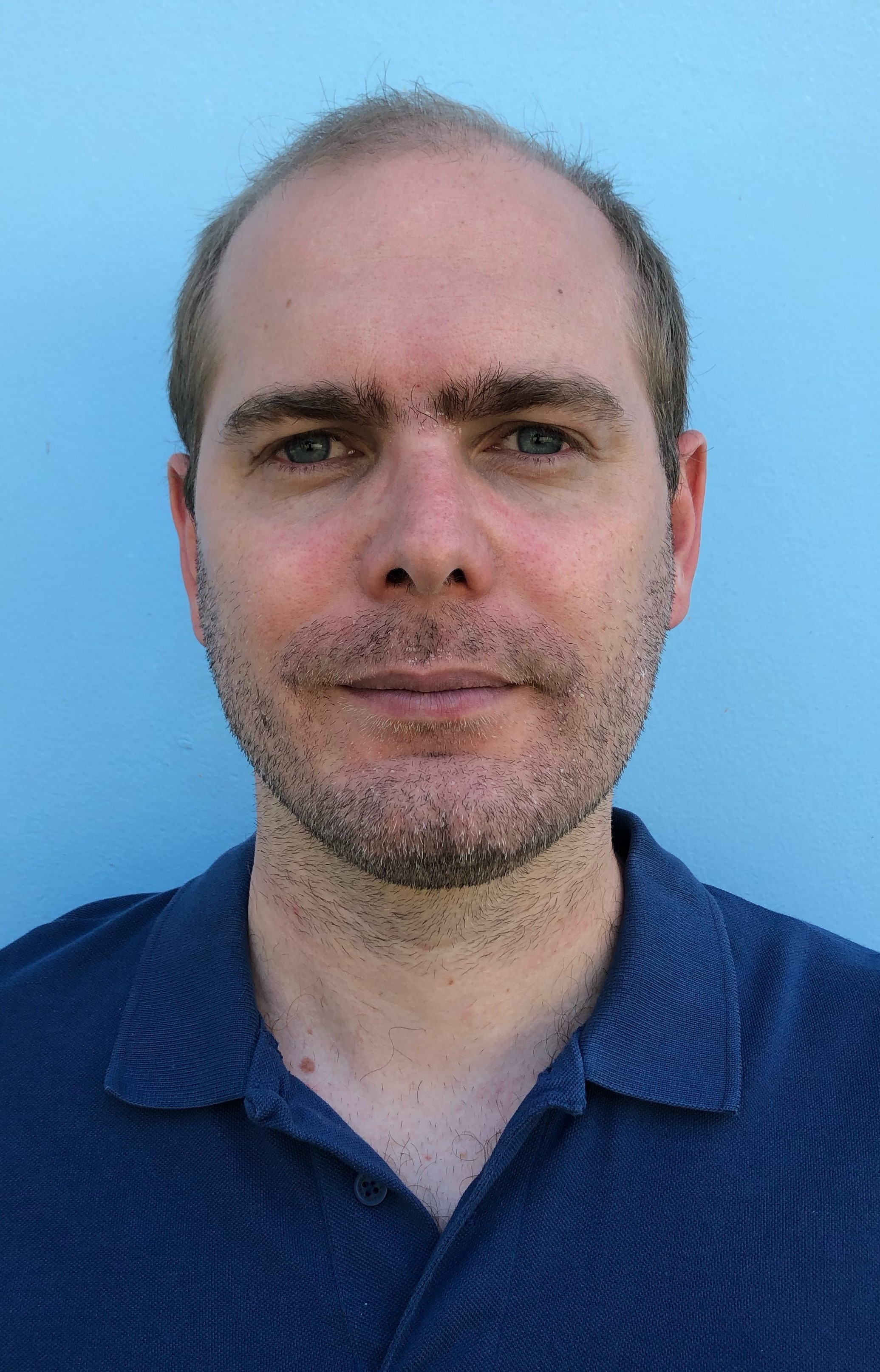}}]{\bf{Marcos Schwarz} -} is Research \& Development Manager at the Brazilian Research and Education Network (RNP). Since 2014, has been working with R\&D teams and projects in Cyber-infraestructures, involving advanced internet, network performance monitoring, dynamic circuit networks, and orchestration in SDN and Cloud-Native. He holds a B.Sc. degree in Computer Science from Santa Catarina State University (UDESC) in 2011 and a M.S. degree in Computer Engineering from the University of São Paulo (USP) in 2014.
\end{IEEEbiography}

\vskip -2\baselineskip plus -1fil

\begin{IEEEbiography}
    [{\includegraphics[width=1in,height=1.25in,clip,keepaspectratio]{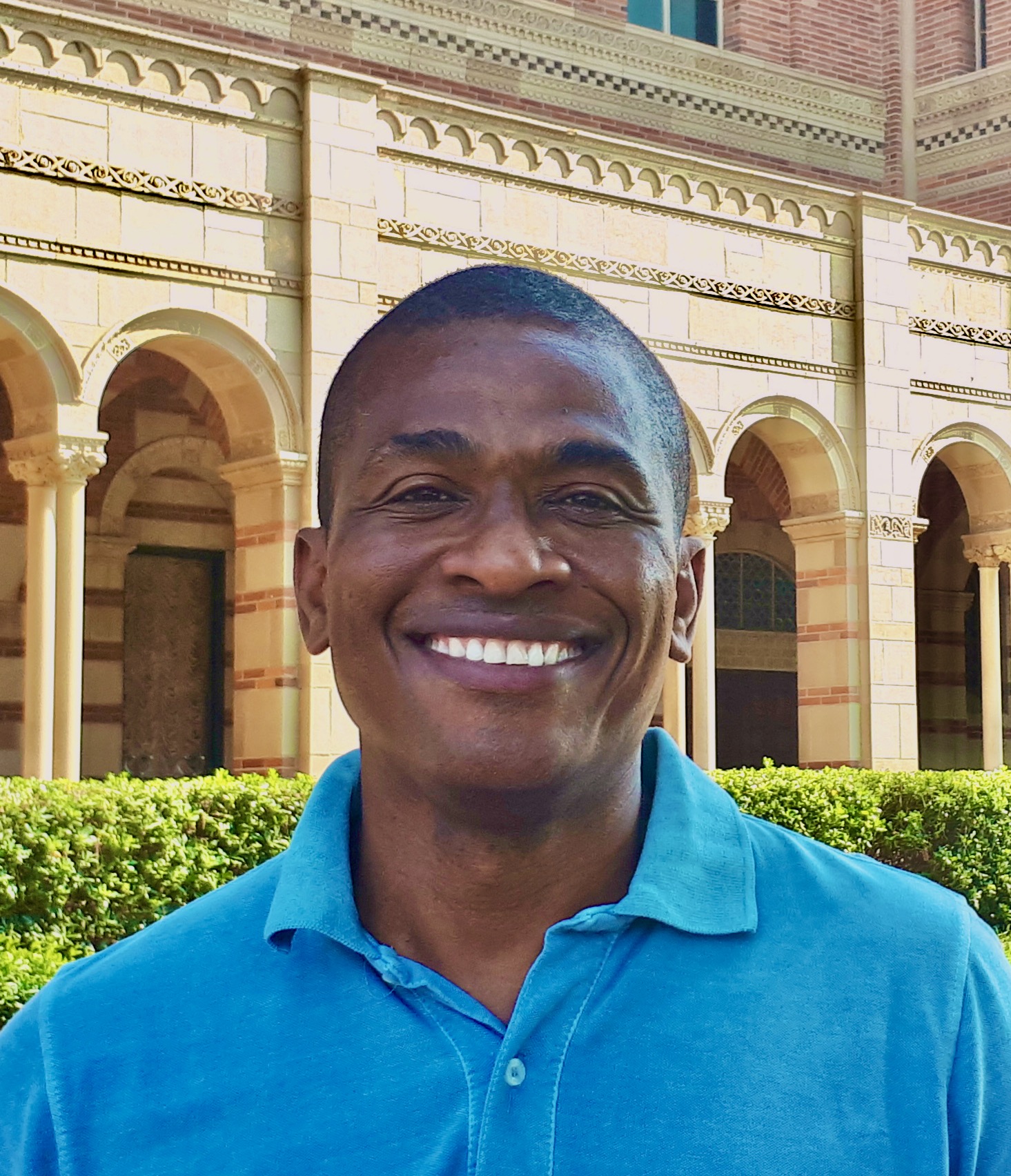}}]{\bf{LEOBINO N. SAMPAIO} - (Member, IEEE)} an Associate Professor in the Computing Department (DCC) in Federal University of Bahia (UFBA). He holds a Ph.D. in computer science from Federal University of Pernambuco (UFPE) and was a visiting researcher with the Computer Science Department of the University of California, Los Angeles (UCLA) in the United States. His current research interests include future Internet architectures and network performance evaluation.

\end{IEEEbiography}

\vskip -2\baselineskip plus -1fil

\begin{IEEEbiography}
    [{\includegraphics[width=1in,height=1.25in,clip,keepaspectratio]{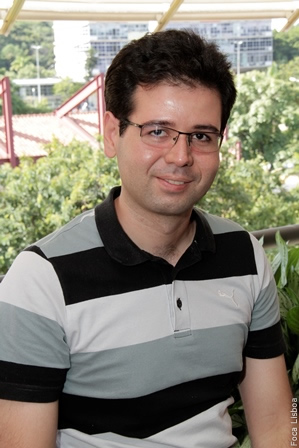}}]{\bf{Daniel F. Macedo} -}  an Associate Professor in the Computing Department (DCC) in Federal University of Minas Gerais (UFMG), Brazil. He has productivity in research scholarship (Bolsa PQ – CNPq) level 2. He was a post-doc researcher at UFMG, Brazil. He holds a Ph.D. in computer science from Université Pierre et Marie Curie-ParisVI (2009). He also holds an M.Sc. and a B.Sc. in Computer Science from the Federal University of Minas Gerais (2006). His research interests are wireless networks,  network management, and autonomic networks.
\end{IEEEbiography}

\vskip -2\baselineskip plus -1fil

\begin{IEEEbiography}
    [{\includegraphics[width=1in,height=1.25in,clip,keepaspectratio]{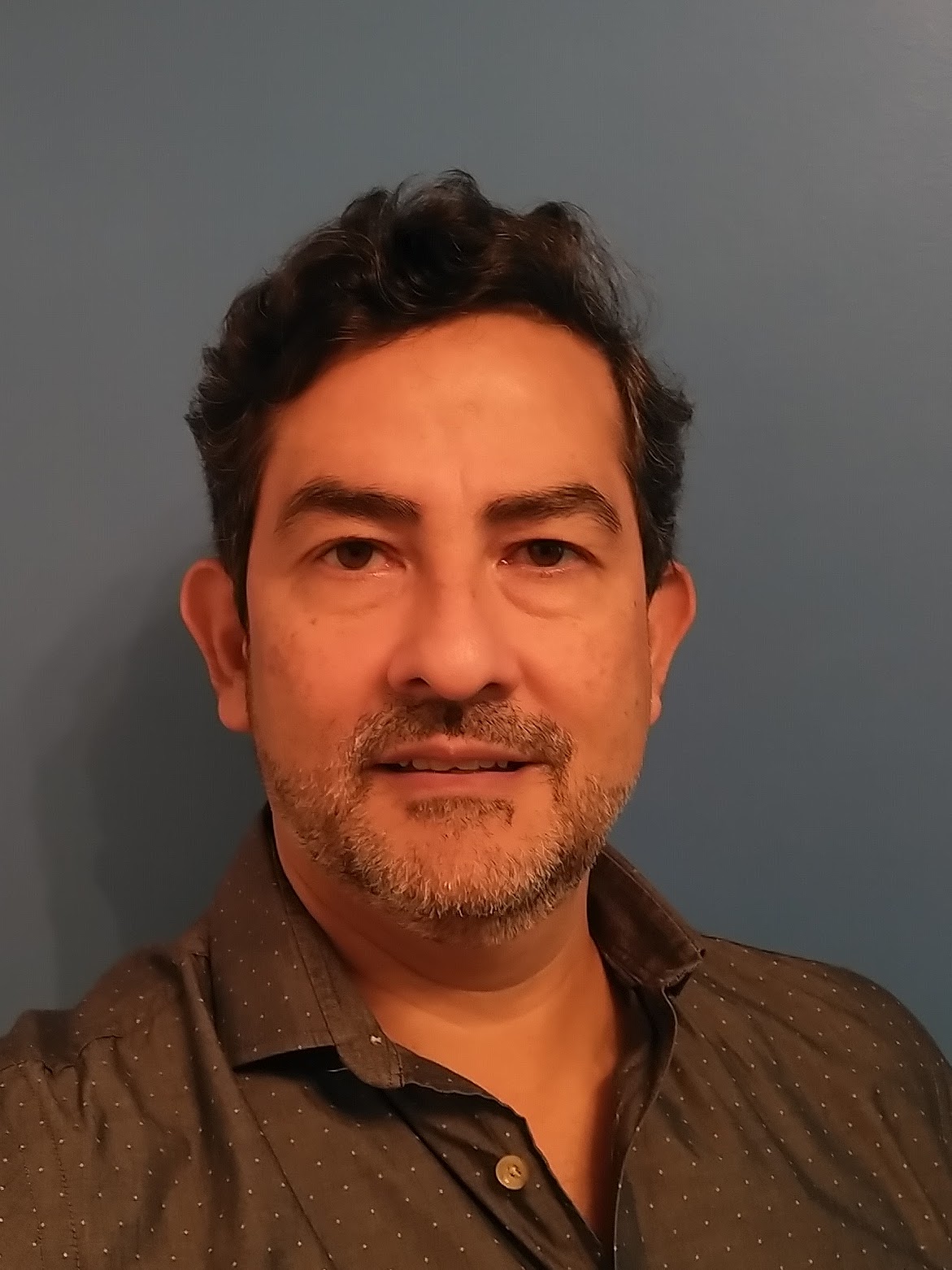}}]{\bf{José F. de Rezende} -} received the Ph.D. degree in computer science from Université Pierre et Marie Curie, in 1997. He was an Associate Researcher with Université Pierre et Marie Curie, in 1997. Since 1998, he has been an Associate Professor with the Federal University of Rio de Janeiro. His research interests include distributed multimedia applications, QoS in the internet, mobile networks, wireless communication, and experimental platforms.

\end{IEEEbiography}

\vskip -2\baselineskip plus -1fil

\begin{IEEEbiography}[{\includegraphics[width=1in,height=1.25in,clip]{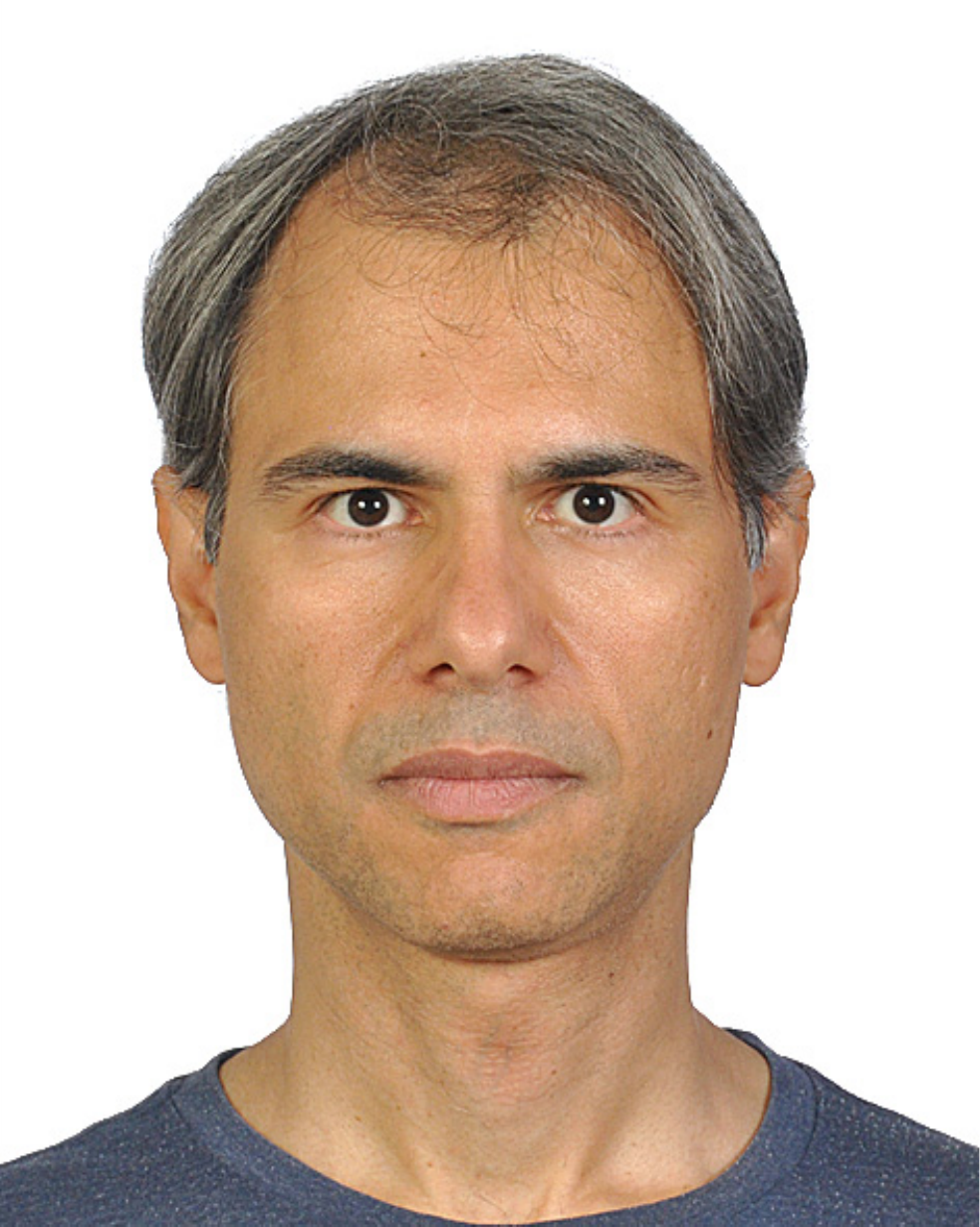}}]{\bf{KLEBER V. CARDOSO} -}is an associate professor at the Institute of Informatics - Universidade Federal de Goi\'{a}s (UFG), where he has been a professor and researcher since 2009. He holds a degree in Computer Science from Universidade Federal de Goi\'{a}s (1997), has MSc (2002) and Ph.D. (2009) in Electrical Engineering from COPPE - Universidade Federal do Rio de Janeiro. In 2015, he spent his sabbatical at Virginia Tech (in the USA) and, in 2020, at Inria Saclay Research Center (in France). His research is focused on the following topics: wireless networks, SDN, virtualization, resource allocation, and performance evaluation.
\end{IEEEbiography}

\vskip -2\baselineskip plus -1fil

\begin{IEEEbiography}
    [{\includegraphics[width=1in,height=1.25in,clip,keepaspectratio]{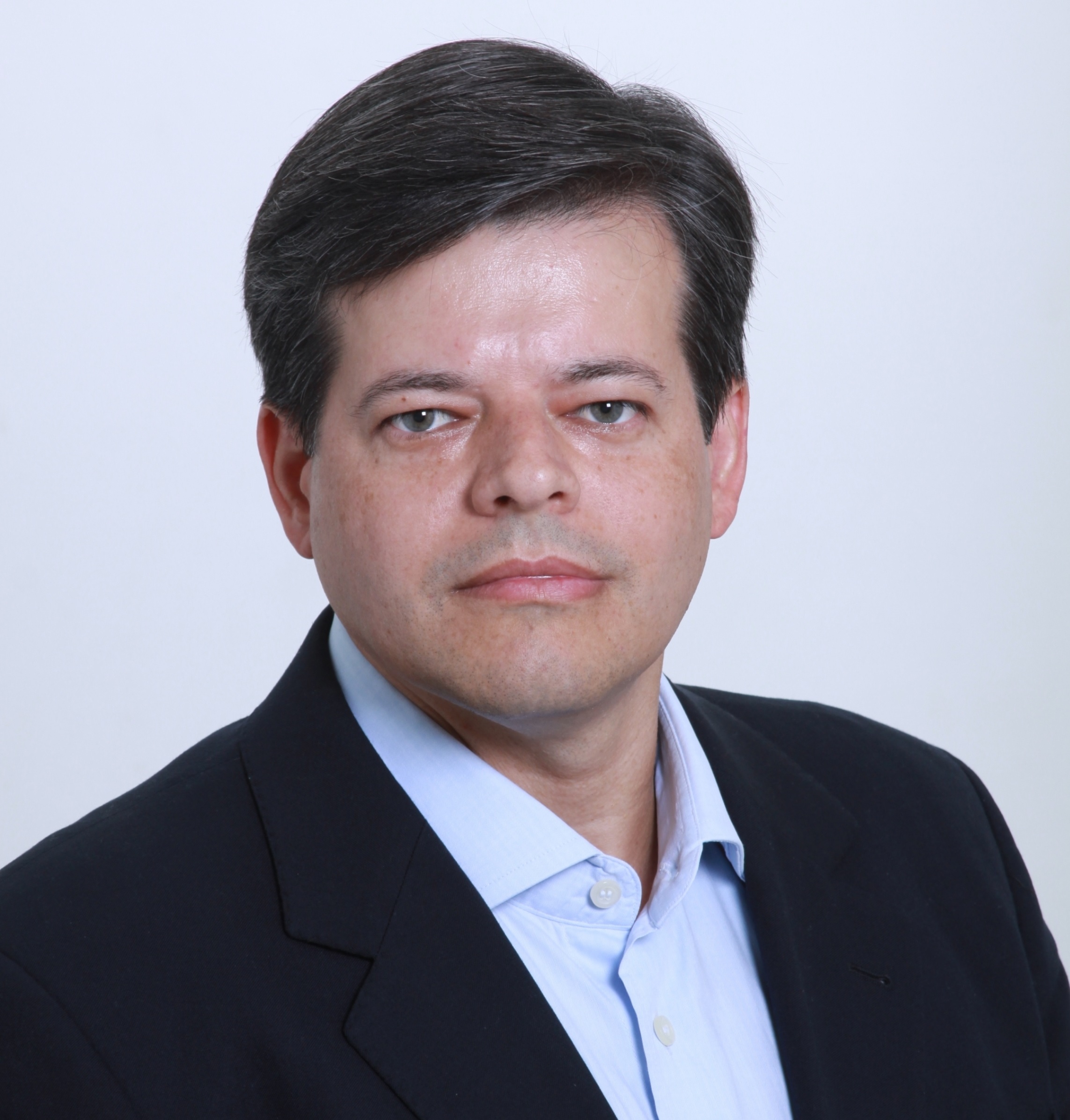}}]{\bf{FLÁVIO DE OLIVEIRA SILVA}  (Member, IEEE) -}
    is a professor in the Faculty of Computing at the Federal University of Uberlândia (UFU) and received a Ph.D. in 2013 from the University of São Paulo. He is a member of ACM, IEEE, and SBC and has had several papers published and presented at conferences worldwide. He is a reviewer for several journals and a member of TCPs at several IEEE conferences. His main research interests are future networks, IoT, network softwarization (SDN and NFV), future intelligent applications and systems, cloud computing, and software-based innovation.
\end{IEEEbiography}

\EOD

\end{document}

%% file: Acronym/acronym.tex

\begin{acronym}

\acro{3GPP}{3rd Generation Partnership Project}
\acro{5GnB}[5G\&B]{5G \& Beyond}
\acro{AI}{Artificial Intelligence} 
\acro{API}{Application Programming Interface}
\acro{CNN}{Convolutional Neural Network}
\acro{DDoS}{Distributed Denial of Service}
\acro{DQN}{Deep Q-Network}
\acro{DNN}{Deep Neural Network} 
\acro{DNNs}{Deep Neural Networks}
\acro{DRL}[DRL]{Deep Reinforcement Learning}
\acro{DNN}{Deep Neural Network} 
\acro{E2E}[E2E]{end-to-end}
\acro{ESG}{Environmental, Social, and Governance}
\acro{ETSI}{European Telecommunications Standards Institute}
\acro{eMBB}{Enhanced Mobile Broadband}
\acro{FL}{Federated Learning}
\acro{IAM}{Identity and Access Management}
\acro{ICT}{Information and Communications Technology}
\acro{IIoT}{Industrial IoT}
\acro{InP}{Infrastructure Provider}
\acro{IoT}{Internet of Things}
\acro{IT}{Information Technology}
\acro{KNN}{K-Nearest Neighbors}
\acro{KPI}{Key Performance Indicator}
\acro{KQI}{Key Quality Indicator}
\acro{LSTM}{Long Short-Term Memory}
\acro{MEC}{Multi-Access Edge Computing}
\acro{ML}{Machine Learning}
\acro{MLaaS}{Machine Learning as a Service}
\acro{MLP}{Multilayer Perceptron}
\acro{MSE}{Mean Squared Error}
\acro{MVNO}{Mobile Virtual Network Operator}
\acro{NFV}{Network Virtualization Function}
\acro{NS}{Network Slicing}
\acro{NGMN}{Next Generation Mobile Networks}
\acro{PT}{Prospect Theory}
\acro{PUE}{Power Usage Effectiveness}
\acro{QoE}{Quality of Experience}
\acro{QoT}{Quality of Transmission}
\acro{SFC}{service function chain}
\acro{SFI2}{Slicing Future Internet Infrastructures}
\acro{UAV}{Unmanned Aerial Vehicle}
\acro{UE}{User Equipment}
\acro{URLLC}{Ultra-Reliable and Low-Latency Communications}
\acro{RAN}{Radio Access Network}
\acro{RL}{Reinforcement Learning}
\acro{SLA}{Service Level Agreement}
\acro{SlaaS}{Slice-as-a-Service}
\acro{SVM}{Support Vector Machine}
\acro{V2I}{Vehicle-to-Infrastructure}
\acro{V2N}{Vehicle-to-Network}
\acro{V2P}{Vehicle-to-Pedestrian}
\acro{V2V}{Vehicle-to-Vehicle}
\acro{V2X}{Vehicle-to-Everything}
\acro{VM}{Virtual Machine}
\acro{VNE}{Virtual Network Embedding}
\acro{VNEP}[VNEP]{Virtual Network Embedding Problem}
\acro{VNF}[VNF]{Virtualized Network Function}

\end{acronym}